\documentclass[nofootinbib,a4paper,12pt]{article}
\usepackage{jheppub}
 \usepackage{etoolbox}                    
 \makeatletter
 \patchcmd{\maketitle}{\@fpheader}{Prepared for submission to JHEP}{}{}
 \makeatother
 
\usepackage[T1]{fontenc}
\usepackage[utf8]{inputenc}
\usepackage{float}
\usepackage{slashed}
\usepackage{multicol,multirow}
\usepackage{amssymb}
\usepackage{amsmath}
\usepackage{subcaption}
\usepackage{array}
\usepackage{color}
\usepackage{hyperref}
\hypersetup{colorlinks=true}
\usepackage{longtable}
\usepackage{graphics,graphicx,epsfig,ulem,booktabs}
\usepackage{hhline}
\numberwithin{equation}{section}

\definecolor{green}{rgb}{0.1,0.85,0.2}
\definecolor{darkgreen}{rgb}{0.08,0.7,0.15}
\definecolor{orange}{rgb}{0.95,0.5,0.2}
\definecolor{cyan}{rgb}{0.0,0.75,0.8}

\newcolumntype{C}[1]{>{\centering\let\newline\\\arraybackslash\hspace{0pt}}m{#1}}

\newcommand{\GeV}{\,\textrm{GeV}}

\newcommand{\psin}{{\rm ps^{-1}}}

\newcommand{\mupi}[1]{\mu_\pi^2 (#1)}
\newcommand{\muG}[1]{\mu_G^2 (#1)}
\newcommand{\rhoD}[1]{\rho_D^3 (#1)}

\newcommand{\Lab}{\Lambda_b^0}
\newcommand{\Xibo}{\Xi_b^0}
\newcommand{\Xibm}{\Xi_b^-}
\newcommand{\Omb}{\Omega_b^-}
\newcommand{\Lac}{\Lambda_c^+}

\newcommand{\Omc}{\Omega_c^0}
\newcommand{\Bary}{\mathcal{B}}
\newcommand{\bary}{b}

\newcommand{\mes}{m}

\newcommand{\Opsixprime}[2]{\mathcal{O}_{#1}^{#2}}

\newcommand{\Opsixtprime}[2]{\tilde{\mathcal{O}}_{#1}^{#2}}

\newcommand{\intm}{\textrm{int}^{-}}
\newcommand{\intp}{\textrm{int}^{+}}
\newcommand{\exc}{\textrm{exc}}

\title{Quark-hadron duality at work:
\\
lifetimes of bottom baryons}

\preprint{RBI-ThPhys-2022-48 \\ \hspace*{\fill} SI-HEP-2022-38 \\ \hspace*{\fill} SFB-257-P3H-22-126}

\author[a]{James Gratrex,}
\author[b]{Alexander Lenz,}
\author[a]{Bla\v zenka Meli\' c,}
\author[a]{Ivan  Ni\v sand\v zi\'c,}
\author[b]{Maria~Laura~Piscopo,}
\author[b]{Aleksey V. Rusov}

\affiliation[a]{Division of Theoretical Physics, Ru\dj er Bo\v skovi\'c Institute, Bijeni\v cka cesta 54, 10000, Zagreb, Croatia}
\affiliation[b]{Physik Department, Universit\"{a}t Siegen, Walter-Flex-Str. 3, 57068 Siegen, Germany}

\emailAdd{jgratrex@irb.hr}
\emailAdd{alexander.lenz@uni-siegen.de}
\emailAdd{maria.piscopo@uni-siegen.de}
\emailAdd{blazenka.melic@irb.hr}
\emailAdd{ivan.nisandzic@irb.hr}
\emailAdd{rusov@physik.uni-siegen.de}

\abstract{In the 1990s, very low experimental values for the lifetime ratio $\tau (\Lambda_b)/\tau(B_d)$ triggered a considerable amount of doubt in the applicability of the heavy quark expansion~(HQE), 
which is based on the assumption of quark-hadron duality (QHD) for inclusive total decay rates. However, these low values turned out to be the result of purely experimental problems, and the current HFLAV average reads $\tau (\Lambda_b)/\tau(B_d) = 0.969(6)$. In this work, we present the Standard Model
predictions for the $b$-baryon lifetimes within the framework of the HQE. In particular, we include for the first time the contribution of the Darwin term and we update the estimates for the matrix elements of the dimension-six four-quark operators.
Within experimental and theoretical uncertainties, we find excellent agreement between the data and the HQE predictions, and thus no indication for any visible violation of QHD. Our numerical results can be summarised by the ratios
$\tau (\Lambda_b)/\tau(B_d) = 0.955(14)$,
$\tau (\Omb)/\tau(B_d) = 1.081(42)$,
and
$\tau (\Xibo)/\tau (\Xibm) = 0.929(28)$.
}

\begin{document}

\maketitle

\section{Introduction}
\begin{figure}[th]
    \centering
    \includegraphics[scale=1.0]{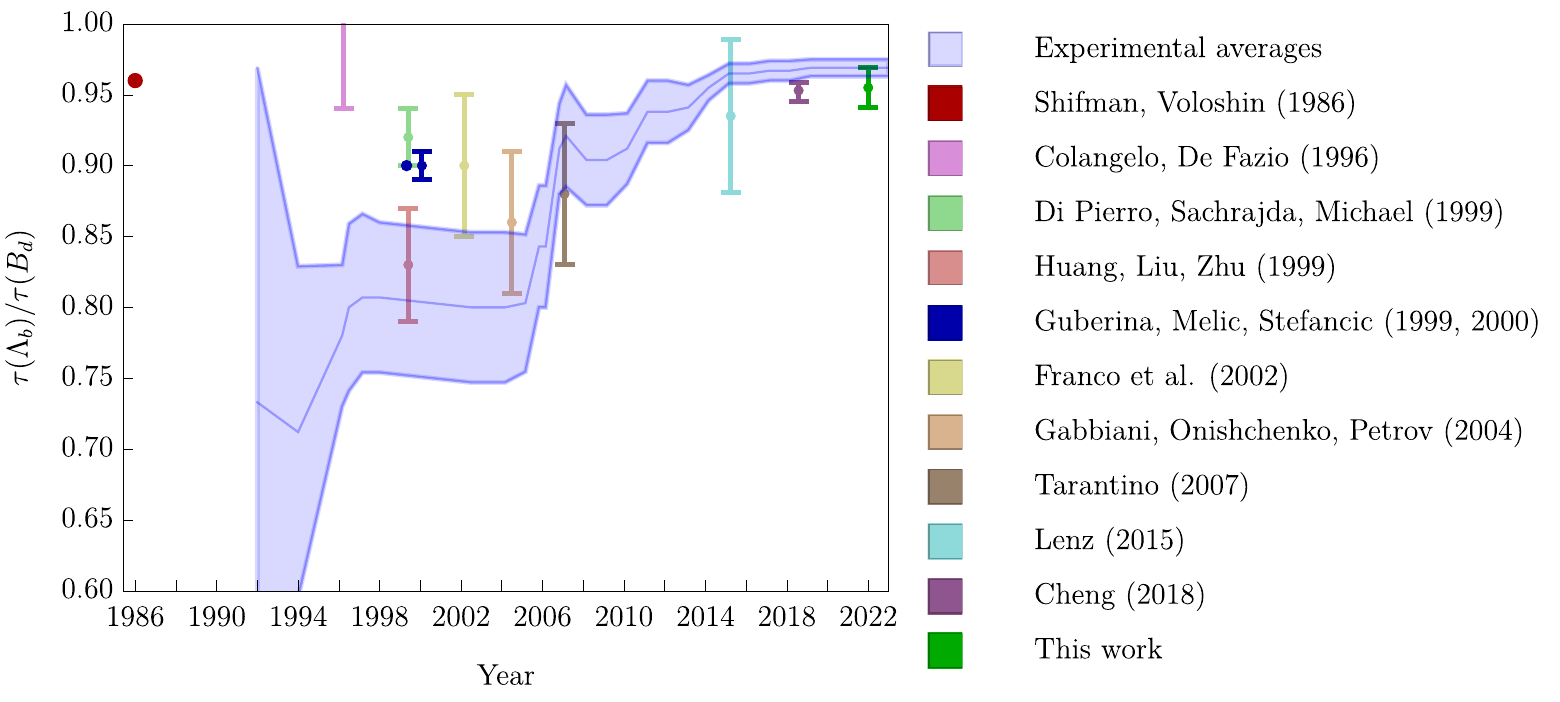}
    \caption{History of the lifetime ratio $\tau (\Lambda_b)/\tau(B_d)$: experiment (lilac) vs. selected  theory predictions:
     {\it Shifman, Voloshin} (1986) \cite{Shifman:1986mx}, {\it Colangelo, De Fazio} (1996) \cite{Colangelo:1996ta}, {\it Di Pierro, Sachrajda, Michael} (1999) \cite{DiPierro:1999tb}, {\it Huang, Liu, Zhu} (1999) \cite{Huang:1999xj}, {\it Guberina, Melic, Stefancic} (1999, 2000) \cite{Guberina:1999bw, Guberina:1999yj}, {\it Franco et al} (2002) \cite{Franco:2002fc}, {\it Gabbiani, Onishchenko, Petrov} (2004) \cite{Gabbiani:2004tp}, {\it Tarantino} (2007) \cite{Tarantino:2007nf}, {\it Lenz} (2015) \cite{Lenz:2014jha}, {\it Cheng} (2018) \cite{Cheng:2018rkz}, and this work.
} 
    \label{fig:tau-Lambda-b-to-tau-Bd-history-Exp-vs-Th}
\end{figure}

Lifetimes are among the most fundamental properties of particles. For weakly decaying hadrons containing a heavy $b$-quark, the lifetimes can be determined theoretically within the framework of the heavy quark expansion (HQE), whose origin goes back to the 1980s \cite{Shifman:1986mx}; see~\cite{Lenz:2014jha} for a review.
According to the HQE, the total decay rate  of a bottom hadron can be described as an  expansion in inverse powers of the heavy quark mass, i.e.\ in $\Lambda_{\rm QCD} / m_b$, with $\Lambda_{\rm QCD}$ being a typical non-perturbative hadronic scale much smaller than the mass of the $b$-quark. The leading term in this expansion is given by the decay of a free $b$-quark, and is completely independent of the decaying hadron.
Taking only this contribution into account would therefore lead to the expectation of equal lifetimes for different $b$-hadrons. Corrections to this picture, and thus deviations of the lifetime
ratios from one, are suppressed by at least two powers of the $b$-quark mass.
Without knowing the size of higher-order QCD corrections, and with only rough estimates for the matrix elements arising in the HQE, the naive expectation in 1986 \cite{Shifman:1986mx} was
\begin{eqnarray}
\left. \frac{\tau (B^+)}{\tau (B_d)} \right|^{\rm HQE \, 1986}
\approx 1.1
\, ,
\hspace{1cm}
\left. \frac{\tau (B_s)}{\tau (B_d)} \right|^{\rm HQE \, 1986}
\approx 1
\, ,
\hspace{1cm}
\left. \frac{\tau (\Lab)}{\tau (B_d)} \right|^{\rm HQE \, 1986}
\approx 0.96
\, .
\label{eq:HQE_1986}
\end{eqnarray}
Surprisingly, early measurements of the $\Lab$ lifetime resulted in values which were considerably lower than the first theory expectations, as shown in figure~\ref{fig:tau-Lambda-b-to-tau-Bd-history-Exp-vs-Th}.\footnote{The $\Lab$ baryon was discovered in 1991 in proton-antiproton collisions by the UA1 collaboration, based on data taken in 1988/89 \cite{UA1:1991vse}. 
The first measurement of the $\Lambda_b$ lifetime was performed by the ALEPH collaboration in 1992 \cite{ALEPH:1992yid}, based on LEP $e^+ e^-$ data taken in 1990/91, and resulting in $ \tau (\Lab) = \left( 1.12_{-0.29}^{+0.32} (\rm stat.) \pm 0.16 (\rm syst.) \right) $~ps. } 

\noindent
In e.g. 1996, the world average for the $\Lambda_b$ lifetime read \cite{Colangelo:1996ta}
\begin{eqnarray}
\tau (\Lambda_b) = (1.18 \pm  0.07 ) \, \mbox{ps}
\, , 
\end{eqnarray}
which corresponded to a lifetime ratio of
\begin{eqnarray}
\frac{\tau (\Lambda_b)}{\tau (B_d)} = (0.75 \pm  0.05 )
\, , 
\label{eq:badratioprediction}
\end{eqnarray}
when using the 1996 world average for the $B_d$ lifetime \cite{Colangelo:1996ta}.
As these experimental results were more than four standard deviations below the naive expectation in eq.~(\ref{eq:HQE_1986}), a considerable amount of interest was triggered in the theory community, with various efforts made to accommodate the result \eqref{eq:badratioprediction} within the HQE.
In \cite{Huang:1999xj}, the possibility of anomalously large matrix elements of dimension-six four-quark operators in the HQE was suggested, which was, however, in conflict with the results of
\cite{Colangelo:1996ta,Rosner:1996fy,DiPierro:1999tb}; while large contributions from dimension-seven four-quark operators were considered in \cite{Gabbiani:2004tp}.

Separately, the validity of the HQE itself was questioned e.g.\ in~\cite{Altarelli:1996gt,Cheng:1997xba,Ito:1997qq}, with \cite{Altarelli:1996gt,Cheng:1997xba} suggesting a violation of local quark-hadron duality (QHD), see e.g.\ \cite{Jubb:2016mvq} for a brief introduction to the concept of QHD. However, the proposal in \cite{Altarelli:1996gt,Cheng:1997xba} was heavily criticised since it would have required huge $1/m_b$ corrections, which cannot be reconciled with the operator product expansion approach, see e.g.\ \cite{Bigi:1999bq}. 
The notion of QHD was introduced in 1975 by Poggio, Quinn, and Weinberg \cite{Poggio:1975af} to equate the hadronic process 
$e^+ + e^- \to \rm hadrons$ with the quark-level process $e^+ + e^- \to \rm quarks$.
In the case of the total decay rate of a $B$-hadron, we can write 
\begin{equation}
    \Gamma^{\rm tot}(B) =  
    \!\!\!\!\! \sum\limits_{\tiny \mbox{all possible hadrons}} \!\!\!\!\! 
    \Gamma (B \to {\rm hadrons + leptons}) =
    \!\!\!\!\! \sum\limits_{\tiny \mbox{all possible quarks}} \!\!\!\!\! 
    \Gamma (B \to {\rm quarks + leptons}) \,,
\end{equation}
and QHD-violating contributions in the HQE could correspond to non-perturbative  terms such as $\exp[-m_b/\Lambda_{\rm QCD}]$,
e.g.\ \cite{Shifman:2000jv,Bigi:2001ys}. Since an exact proof of QHD would require one to explicitly solve QCD, which is clearly not possible currently, 
we can consider two strategies in order to investigate the possible size of duality-violating effects.
Firstly, one could study simplified models of QCD, like the 't Hooft model, a 1+1 dimension model for QCD, 
e.g.~\cite{tHooft:1974pnl,Bigi:1998kc,Lebed:2000gm,Grinstein:2001zq,Shifman:2000jv,Bigi:2001ys}, or instanton-based and 
resonance-based models, e.g.\ \cite{Chibisov:1996wf,Shifman:2000jv,Bigi:2001ys}. 
Studies of the 't Hooft model indicate the presence of duality-violating terms, albeit at very high orders in the HQE and thus numerically irrelevant. Nevertheless, it is not clear what stringent conclusions can be drawn from the study of a 1+1 dimensional model for the real 3+1 dimensional world. The second approach is purely phenomenological, and is based on comparing experimental data with precise HQE predictions. In the present work, we follow this latter strategy. 

Ultimately, it turned out that the low values for the $\Lambda_b$ lifetime were purely an experimental problem, and the current world average for $\tau (\Lambda_b) / \tau(B_d)$ \cite{Zyla:2020zbs} now agrees perfectly with the estimates from 1986. This is clearly shown in figure~\ref{fig:tau-Lambda-b-to-tau-Bd-history-Exp-vs-Th}, where we plot the time evolution of the experimental measurements for this observable, from 1992 onwards, in comparison to selected theory predictions
\cite{Shifman:1986mx,Colangelo:1996ta,DiPierro:1999tb,Huang:1999xj,Guberina:1999bw, Guberina:1999yj, Franco:2002fc,Gabbiani:2004tp,Tarantino:2007nf,Lenz:2014jha,Cheng:2018rkz}, as well as our result.
Based on the measurements in~\cite{DELPHI:1995jet,OPAL:1995nmi,DELPHI:1996eqs,ALEPH:1996kuy,CDF:1996fsp,ALEPH:1997ake,OPAL:1997ufs,DELPHI:1999con,DELPHI:2005zmk,D0:2007pfp,CDF:2009sse,D0:2012hfl,ATLAS:2012cvl,CMS:2013bcs,LHCb:2014qsd,LHCb:2014wvs,CDF:2014mon,LHCb:2014wqn,LHCb:2014chk,LHCb:2014jst,LHCb:2016coe,CMS:2017ygm},
HFLAV \cite{Amhis:2022mac} quotes for the lifetimes of  different weakly decaying $b$-baryons the precise
values listed in table~\ref{tab:exp-data} and the lifetime ratios listed in table~\ref{tab:exp-data_ratio}.
\begin{table}[h]
\centering
\renewcommand{\arraystretch}{1.6}
    \begin{tabular}{|c||c|c|c|c||c|}
    \hline
         & $\Lab$ 
         & $\Xibo $ 
         & $\Xibm $
         & $\Omb $ 
         & $B_d^0$
         \\
\hhline{|=||=|=|=|=||=|}
    $\tau \, [{\rm ps}]$ 
    & $1.471 \pm 0.009 $ 
    & $1.480 \pm 0.030 $ 
    & $1.572 \pm 0.040 $
    & $1.64^{+0.18}_{-0.17} $
    & $1.519 \pm 0.004$
    \\
    \hline
     $\Gamma \, [{\rm ps}^{-1}]$ 
     & $0.680 \pm 0.004 $ 
     & $0.676 \pm 0.014 $ 
     & $0.636 \pm 0.016 $
     & $0.610^{+0.070}_{-0.066} $
     & $0.636 \pm 0.016 $
    \\
    \hline
    \end{tabular}
    \caption{
    HFLAV averages of the experimental determinations of $b$-baryon lifetimes~\cite{Amhis:2022mac}. We also include the most recent value of the $B_d^0$ meson lifetime, which we use in our predictions for the lifetime ratios with the baryons.
   }
    \label{tab:exp-data}
\end{table}
\begin{table}[h]
\centering
\renewcommand{\arraystretch}{1.6}
    \begin{tabular}{|c||c|}
    \hline
 $\tau(\Lab)/\tau (B_d^0)$ & $0.969 \pm 0.006$
 \\
 \hline
 $\tau (\Xibo)/ \tau (\Xibm) $ & $0.929 \pm 0.028$
\\
    \hline
    \end{tabular}
    \caption{HFLAV averages of the experimental determinations of  $b$-baryon lifetime ratios~\cite{Amhis:2022mac}.}
    \label{tab:exp-data_ratio}
\end{table}

In this paper, we present theory predictions for the lifetimes of baryons containing a heavy $b$-quark, as a continuation of our work on the study of lifetimes of 
$D$ mesons~\cite{King:2021xqp,Gratrex:2022xpm},
$B$ mesons~\cite{Lenz:2022rbq},
and charmed baryons~\cite{Gratrex:2022xpm}.
Besides implementing for the first time the recently determined Wilson coefficient of the Darwin operator~\cite{Mannel:2020fts,Lenz:2020oce,Moreno:2020rmk,Piscopo:2021ogu}, we include radiative QCD corrections 
to the Wilson coefficients, where available, and update all the relevant numerical inputs, including new estimates for the non-perturbative matrix elements. 
We present predictions for the decay rates of the $\Lab, \,\Xibo,\, \Xibm,$ and $\Omb$ baryons, and their lifetime ratios, as well as lifetime ratios of these baryons with the $B_d^0$ meson. Within uncertainties, our results are in excellent agreement with the experimental data. 
Moreover, we give predictions for the inclusive $b$-baryon semileptonic branching fractions, although in this case there are no current experimental determinations. 

The remainder of the paper is structured as follows. In section~\ref{sec:Th-fram} we present the theoretical framework. Specifically, in section~\ref{sec:Eff-Ham}, we briefly describe the structure of the HQE, followed by the discussion of short-distance contributions in section~\ref{sec:short-dist-contr}, and the analysis of non-perturbative matrix elements in section~\ref{sec:matrixelements}. Section~\ref{sec:Num-analysis} contains the description of the numerical analysis and our predictions for the $b$-baryon lifetimes, lifetime ratios, and semileptonic branching fractions. We conclude in section~\ref{sec:conclusions}. Appendix~\ref{app:1} contains numerical values of the input parameters used in the analysis, while in appendix~\ref{app:2}, we provide the analytical expressions at LO-QCD for the dimension-six four-quark operator contributions. 

\section{Theoretical framework}
\label{sec:Th-fram}
 \subsection{Effective Hamiltonian and HQE}
\label{sec:Eff-Ham}
Weak $b$-quark decays can be described by the effective Hamiltonian~\cite{Buchalla:1995vs} 
\begin{equation}
  {\cal H}_{\rm eff}  =   {\cal H}_{\rm eff}^{\rm NL} + {\cal H}_{\rm eff}^{\rm SL} + {\cal H}_{\rm eff}^{\rm rare} \,.
\label{eq:Heff-complete}
\end{equation}
In the above equation, ${\cal H}_{\rm eff}^{\rm NL}$ parametrises the contribution of non-leptonic $b$-quark transitions 
\begin{align}
  {\cal H}_{\rm eff}^{\rm NL} = 
  \frac{G_F}{\sqrt{2}} \sum_{q_3 = d, s}
  \left[\,
   \sum_{\substack{q_{1,2} = u, c} } \!\! \lambda_{q_1 q_2 q_3} 
  \Bigl(C_1 (\mu_1) \, Q_1^{q_1 q_2 q_3}  + C_2 (\mu_1) \, Q_2^{q_1 q_2 q_3}  \Bigr)
    -  \lambda_{q_3} 
  \!\! \!\! \sum \limits_{j=3, \ldots, 6, 8} \! \!\! C_j (\mu_1) \, Q_j^{q_3} 
   \right] + {\rm h.c.}\, ,
   \label{eq:Heff-NL}
\end{align}
where $\lambda_{q_1 q_2 q_3} = V_{q_1 b}^* V_{q_2 q_3} $
and $\lambda_{q_3} = V_{tb}^* V_{tq_3} $ stand for the corresponding CKM factors, 
$C_i (\mu_1)$ denote the Wilson coefficients of the $\Delta B = 1$ effective operators evaluated at the renormalisation scale $\mu_1 \sim m_b$. 
$Q_{1,2}^{q_1 q_2 q_3}$ and $Q_j^{q_3}$, with $j = 3, \ldots, 6,$ and $Q_8^q$, respectively denote the current-current,\footnote{Note that $Q_1^{q_1q_2 q_3}$ in our notation is the colour-singlet operator, following~\cite{King:2021xqp,Lenz:2022rbq} and
contrary to e.g.~\cite{Buchalla:1995vs,Gratrex:2022xpm}.} penguin, and chromomagnetic operators, and are explicitly
\begin{equation}
Q_1^{q_1 q_2 q_3} 
 =   
\left(\bar b^i \, \Gamma_\mu \, q_1^i \right)
\left(\bar q_2^j \, \Gamma^\mu  \, q_3^j \right)\,,
\qquad 
Q_2^{q_1 q_2 q_3} 
 =  
\left(\bar b^i  \, \Gamma_\mu  \, q_1^j \right)
\left(\bar{q}_2^j \, \Gamma^\mu  \, q_3^i \right)\,,
\label{eq:Q12}
\end{equation}
\begin{align}
Q_3^{q_3} 
& 
= (\bar b^i \, \Gamma_\mu \, q_3^i) \sum_{q} ( \bar q^j \, \Gamma^\mu \, q^j)
\,, \qquad 
Q_4^{q_3} = (\bar b^i \, \Gamma_\mu \, q_3^j) \sum_{q} (\bar q^j \, \Gamma^\mu \, q^i)\,, 
\nonumber
\\
Q_5^{q_3} 
& 
=  (\bar b^i \, \Gamma_\mu \, q_3^i) \sum_{q} (\bar q^j \, \Gamma_+^\mu \, q^j)\,, 
\qquad
Q_6^{q_3} = (\bar b^i \, \Gamma_\mu \, q_3^j) \sum_{q} 
(\bar q^j \, \Gamma_+^\mu \, q^i)\,,
\label{eq:Q3456}
\end{align}
\begin{equation}
Q_8^{q_3} = \frac{g_s}{8 \pi^2} m_b
\left(\bar b^i \, \sigma^{\mu\nu} (1 - \gamma_5) t^a_{ij} \, q_3^j \right) G^a_{\mu \nu}\,,
\label{eq:Q8}
\end{equation}
with $\Gamma_\mu = \gamma_\mu(1-\gamma_5)$,  $\Gamma_+^\mu = \gamma^\mu(1+\gamma_5)$, and $\sigma_{\mu \nu} =(i/2) [\gamma_\mu, \gamma_\nu]$, while $i,j = 1, 2, 3,$ are SU(3)$_c$ indices for the quark fields. Moreover, in eq.~\eqref{eq:Q8}, $g_s$~denotes the strong coupling, and $G_{\mu\nu} = G^a_{\mu\nu}t^a $ for $a = 1, \ldots, 8$ is the gluon field strength tensor.
A comparison of the values of the Wilson coefficients for different choices of the scale $\mu_1$ at LO- and NLO-QCD \cite{Buchalla:1995vs} is shown in table~\ref{tab:WCs} in appendix~\ref{app:1}.

The second term in eq.~\eqref{eq:Heff-complete} describes the contribution to the effective Hamiltonian due to semileptonic $b$-quark decays, i.e.
\begin{equation}
{\cal H}_{\rm eff}^{\rm SL} 
= 
\frac{G_F}{\sqrt 2} \sum_{q_1 = u, c \,} \sum_{\, \ell = e, \mu, \tau}
V_{q_1b}^* \, Q^{q_1 \ell} + {\rm h.c.}\,,
\label{eq:Heff-SL}
\end{equation}
with the semileptonic operator
\begin{equation}
    Q^{q_1 \ell} =\left(\bar{b}^i\, \Gamma_\mu \, q_1^i \right)
\left(\bar \nu_\ell \, \Gamma^\mu \, \ell \right)\,.
\end{equation}
Finally, ${\cal H}_{\rm eff}^{\rm rare}$ in eq.~\eqref{eq:Heff-NL} encodes the contribution due to suppressed $b$-quark transitions, which are only relevant for the study of rare decays such as $\Lambda_b \to \Lambda \gamma$ or 
$\Lambda_b \to \Lambda \ell^+ \ell^-$. These modes have very small branching fractions, below the current theoretical sensitivity for lifetimes, and so the effect of ${\cal H}_{\rm eff}^{\rm rare}$ is neglected in this work.

The total decay width of a $b$-baryon ${\cal B}$, with mass~$M_{\cal B}$ and four-momentum $p_{\cal B}$, reads  
\begin{equation}
\Gamma ({\cal B})  = 
\frac{1}{2 M_{\cal B}} \sum_{X}  \int \limits_{\rm PS} (2 \pi)^4  \delta^{(4)}(p_{\cal B}- p_X) \, \,
|\langle X(p_X)| {\cal H}_{\rm eff} | H(p_{\cal B}) \rangle |^2,
\label{eq:Gamma-D}
\end{equation}
where a summation over all possible final states $X$ into which the $b$-baryon can decay has been performed, with PS denoting the corresponding phase space integration. Using the optical theorem, $\Gamma({\cal B})$ can be related to the imaginary part of the forward scattering matrix element of the time-ordered product of the double insertion of the effective Hamiltonian, i.e.
\begin{equation}
\Gamma ({\cal B}) =    \frac{1}{2 M_{\cal B}} {\rm Im}
\langle {\cal B} | {\cal T}| {\cal B} \rangle \, ,
\label{eq:Gamma_opt_th}
\end{equation}
with the transition operator defined as
\begin{equation}
{\cal T}  =  
i \int d^4x 
\,  T \left\{ {\cal H} _{\rm eff} (x) \, ,
 {\cal H} _{\rm eff} (0)  \right\} \, .
 \label{eq:optical_theorem}
\end{equation}
The non-local operator in eq.~\eqref{eq:optical_theorem} can then be evaluated by exploiting the fact that the $b$-quark is heavy, i.e.\ $m_b \gg \Lambda_{\rm QCD}$, where $\Lambda_{\rm QCD}$ defines a typical non-perturbative hadronic scale. In the framework of the HQE
\cite{Khoze:1983yp,Shifman:1984wx,Shifman:1986mx,Bigi:1991ir,Bigi:1992su,Blok:1992hw,Blok:1992he,Bigi:1993ex,Beneke:1998sy,Lenz:2014jha}, the $b$-quark momentum is  decomposed as
\begin{equation}
p_b^\mu = m_b  v^\mu + k^\mu\,,
\label{eq:c-quark-momentum}
\end{equation}
where $v = p_{\cal B}/M_{\cal B}$ is the four-velocity of the $b$-baryon. The residual momentum $k$ in \eqref{eq:c-quark-momentum} accounts for non-perturbative interactions of the $b$-quark with the light degrees of freedom, i.e.\ soft gluons and quarks, inside the hadron, so $k \sim \Lambda_{\rm QCD}$. 
Moreover, the heavy $b$-quark field is parametrised as  
\begin{equation}
b (x) = e^{ - i m_b v \cdot x} b_v (x)\,,  
\label{eq:phase-redef}    
\end{equation}
by factoring out the large component of its momentum and introducing a rescaled field $b_v(x)$, which contains only low oscillation frequencies of order $k$. This field  satisfies
\begin{equation}
    i D_\mu b(x) = e^{- i m_b v \cdot x} (m_b v_\mu + i D_\mu) b_v(x)\,,
    \label{eq:iDmu-b}
\end{equation}
so that the action of the covariant derivative $D_\mu = \partial_\mu - i g_s A_\mu^a \, t^a $ also contains a large contribution proportional to the heavy quark mass alongside a residual term of order $\Lambda_{\rm QCD}$. The rescaled field $b_v(x)$ is related to the heavy quark effective theory (HQET) field $h_v(x)$, see e.g.~\cite{Neubert:1993mb}, by 
\begin{equation}
b_v (x) = h_v (x) + \frac{i \slashed D_\perp}{2 m_b} h_v (x)  
+ {\cal O} \left(\frac{1}{m_b^2} \right)\,,
\label{eq:bv-hv-relation}
\end{equation}
with $D_\perp^\mu = D^\mu - (v \cdot D) \, v^\mu$.
Finally, taking into account eqs.~\eqref{eq:c-quark-momentum}-\eqref{eq:iDmu-b}, the total decay width in eq.~\eqref{eq:Gamma_opt_th} can be systematically expanded in inverse powers of the heavy $b$-quark mass, 
leading to the HQE series, which schematically reads
\begin{equation}
\Gamma({\cal B}) = 
\Gamma_3  +
\Gamma_5 \frac{\langle {\cal O}_5 \rangle}{m_b^2} + 
\Gamma_6 \frac{\langle {\cal O}_6 \rangle}{m_b^3} + ...  
 + 16 \pi^2 
\left( 
  \tilde {\Gamma}_6 \frac{\langle 
  \tilde {\mathcal{O}}_6 \rangle}{m_b^3} 
+ \tilde \Gamma_7 
\frac{\langle \tilde{\mathcal{O}}_7 \rangle}{m_b^4} + ... 
\right).
\label{eq:HQE}
\end{equation}
Here, the $\Gamma_d$ are short-distance functions, which can be computed perturbatively in QCD, i.e.
\begin{equation}
\Gamma_d = \Gamma_d^{(0)} + \frac{\alpha_s}{4 \pi} \Gamma_d^{(1)} 
+ \left(\frac{\alpha_s}{4 \pi}\right)^2 \Gamma_d^{(2)}+ \ldots \, ,  
\label{eq:Gamma-i-pert-series}
\end{equation}
while $\langle {\cal O}_d \rangle \equiv
\langle {\cal B}| {\cal O}_d |{\cal B}  \rangle/(2 M_{\cal B})$ denote the matrix elements of the corresponding $\Delta B = 0$ operators 
${\cal O}_d$ of dimension $d$ in the effective theory. Note that, starting from order $1/m_b^3$, both two- and four-quark operator contributions appear. The latter originate from loop-enhanced diagrams, as reflected by the explicit factor of $16 \pi^2$ in eq.~\eqref{eq:HQE}, and, to avoid confusion in the notation, we use a tilde to label them. 

\subsection{Short-distance contributions}
\label{sec:short-dist-contr}
In this section, we give a brief summary of the short-distance
contributions, cf.\ eqs.~(\ref{eq:HQE}, \ref{eq:Gamma-i-pert-series}), included in our analysis. For more details we refer to the recent studies \cite{King:2021xqp, Gratrex:2022xpm,Lenz:2022rbq}.\footnote{There are some differences in the structure of the HQE for charmed hadrons \cite{King:2021xqp, Gratrex:2022xpm} as opposed to the $b$ sector; see also \cite{Fael:2019umf, Mannel:2021uoz}.}

The coefficients $\Gamma_d, \tilde \Gamma_d$ are analytic functions of the masses of the internal fermions running in the loops. In our analysis, we only include the contribution of the charm-quark and tau-lepton masses, expressed in terms of the two dimensionless parameters
\begin{equation}
    x_c = \frac{m_c^2}{m_b^2} \,, \qquad x_\tau = \frac{m_\tau^2}{m_b^2} \,.
    \label{eq:xcandxtau}
\end{equation} 
As $m_s^2/m_b^2 \approx m_\mu^2/m_b^2 \sim 0.05 \%$, the effect of non-vanishing strange-quark and muon masses is far below the current theoretical accuracy, and hence can be safely neglected.\footnote{ However, we do include strange quark mass corrections in the non-perturbative input, where these effects are much more pronounced, in order to account for SU(3)$_F$-breaking.} 
The leading contribution to the $b$-baryon total width, $\Gamma_3$ in eq.~\eqref{eq:HQE}, is obtained by computing the free $b$-quark decay, and can be compactly expressed as 
\begin{equation}
\Gamma_3 = \Gamma_0 \, c_3 = \Gamma_0 \left( c_3^{(0)} + \frac{\alpha_s}{4 \pi} c_3^{(1)} + \ldots \right)\,,
\end{equation}
where
\begin{equation}
\Gamma_0 = \frac{G_F^2 \, m_b^5}{192 \pi^3} 
|V_{cb}|^2\,,
\label{eq:Gamma0}
\end{equation}
and 
\begin{equation}
    c_3 = 
     {\cal C}_{3, \rm SL}  +
    3 \, C_1^2    \, {\cal C}_{3,11} 
 +  2 \, C_1 C_2  \, {\cal C}_{3,12} 
 +  3 \, C_2^2    \, {\cal C}_{3,22}
 + C_i \, C_j \, {\cal C}^{P}_{3, ij} 
 \,.
 \label{eq:c3-decomposition}
\end{equation}
Above, a summation over all possible non-leptonic and semileptonic modes of the $b$-quark is implicitly assumed, and we have denoted by ${\cal C}_{3, ij}^P$, with $i = 1, 2,$ and $j = 3, \ldots, 6, 8,$ the contribution due to the mixed insertion of the current-current and penguin or chromomagnetic operators. For semileptonic modes, $\alpha_s^3$-corrections have been computed \cite{Fael:2020tow,Czakon:2021ybq}; however, as the accuracy for non-leptonic modes reaches only NLO-QCD, we perform our analysis consistently at this order and do not include the new results for ${\cal C}_{3, {\rm SL}}$. Moreover, following a common counting adopted in the literature \cite{Lenz:1997aa, Lenz:1998qp}, the contribution of the penguin and chromomagnetic operators is treated as a next-to-leading order effect, i.e. ${\cal C}^P_{3, ij} = 0$ at LO-QCD, owing to the small size of the corresponding Wilson coefficients. The result for $c_3$ at LO can be found e.g.\ in~\cite{Lenz:2020oce,Gratrex:2022xpm}. As for the NLO corrections, the analytical expressions for ${\cal C}_{3, 11}$, ${\cal C}_{3, 22}$, and ${\cal C}_{3, {\rm SL}}$ can be extracted from \cite{Hokim:1983yt}, where the computation has been performed for three different final state masses, 
while those for ${\cal C}_{3, 12}$ are derived from the results presented in \cite{Krinner:2013cja} in the case of the $b \to c \bar c s$ transition, and in \cite{Bagan:1994zd} for the remaining modes. Finally the results for ${\cal C}_{3, ij}^P$ are taken from \cite{Krinner:2013cja}.

Power corrections due to two-quark operators are obtained by including the effect of soft gluons as well as  the $1/m_b$-expansion of lower-dimensional matrix elements. At order $1/m_b^2$, the corresponding contribution can be schematically written as 
\begin{equation}
    \Gamma_5 \frac{\langle {\cal O}_5 \rangle}{m_b^2} 
    = \Gamma_0 
    \left[
c_{\pi} \frac{\langle {\cal O}_{\pi}\rangle}{m_b^2} 
    + c_G \, \frac{\langle {\cal O}_{G}\rangle}{m_b^2} 
    \right]\,,
    \label{eq:Gamma_5}
\end{equation}
where the matrix elements of the kinetic and chromomagnetic operators\footnote{Note that, with a little abuse of notation, we refer to both $Q^{q_3}_8$ and ${\cal O}_G$ as chromomagnetic operators. However, as they arise respectively in the $\Delta B = 1$ and $\Delta B = 0$ effective theory, it should be clear from the context to which one we refer.}, given explicitly in eqs.~(\ref{eq:mu-PI}, \ref{eq:mu-G}), are discussed in section \ref{sec:matrixelements}.
In our analysis, again for consistency, we include the coefficients $c_{\pi}$ and $c_{G}$ only at LO-QCD, since $\alpha_s$-corrections have so far been determined only for the semileptonic channels~\cite{Mannel:2015jka}. The coefficient of the kinetic operator is related to that of dimension-three by a purely numerical factor,
$c_{\pi}= - c_3^{(0)}/2$, while the coefficient $c_G$ can be decomposed as
\begin{equation}
    c_G = 
     {\cal C}_{G,\rm SL}+
    3 \, C_1^2    \, {\cal C}_{G,11} 
 +  2 \, C_1 C_2  \, {\cal C}_{G,12} 
 +  3 \, C_2^2    \, {\cal C}_{G,22} \, ,
\label{eq:CG}
\end{equation}
where again a summation over all possible $b$-quark modes is implied.
The expressions for the non-leptonic channels ${\cal C}_{G,ij}$, originally computed in \cite{Blok:1992he, Blok:1992hw, Bigi:1992ne}, can be found e.g. in \cite{Lenz:2020oce,Gratrex:2022xpm}, while
the semileptonic coefficient ${\cal C}_{G, SL}$ is taken from the general result for two different final state masses presented e.g.\ in the appendix of \cite{Mannel:2017jfk}, and first determined in \cite{Balk:1993sz, Falk:1994gw}.
 
At order $1/m_b^3$, both
two- and four-quark operators contribute, cf.\ eq.~\eqref{eq:Gamma-i-pert-series}.
For the former, we can compactly write\footnote{Formally, at dimension-six the basis also includes the spin-orbit operator ${\cal O}_{\rm LS}$. However, by adopting definitions in terms of full covariant derivatives rather than transversal ones, the contribution of ${\cal O}_{\rm LS}$ to the total decay width vanishes. For more detail, see e.g.~\cite{Dassinger:2006md}.}
\begin{eqnarray}
\Gamma_6 \frac{\langle {\cal O}_6 \rangle}{m_b^3}
=
\Gamma_0 \, c_{\rho_D} \frac{\langle {\cal O}_{D}\rangle}{m_b^3} \, ,
\label{eq:Gamma_6}
\end{eqnarray}
where the matrix element of the Darwin operator is defined in eq.~\eqref{eq:rho-D}, while the corresponding short-distance coefficient can be decomposed as
\begin{equation}
    c_{\rho_D} = 
    {\cal C}_{\rho_D,\rm SL} 
    +
    3 \, C_1^2 \, {\cal C}_{\rho_D,11} 
 +  2 \, C_1 C_2 \, {\cal C}_{\rho_D,12} 
 +  3 \, C_2^2 \, {\cal C}_{\rho_D,22}  \,,
\label{eq:crhoD}
\end{equation}
summing again over all $b$-quark decay modes.
As NLO-QCD corrections are only available for semileptonic decays  \cite{Mannel:2019qel, Mannel:2021zzr, Moreno:2022goo}, the accuracy in our analysis again extends to only LO-QCD, identically to the dimension-five contributions.
The complete expressions of ${\cal C}_{\rho_D, ij}$ for all non-leptonic channels have been obtained recently in \cite{Lenz:2020oce, Mannel:2020fts, Moreno:2020rmk,Piscopo:2021ogu}, while the coefficient ${\cal C}_{\rho_D, {SL}}$, first computed in \cite{Gremm:1996df}, can be read off the general results for the case of two different final state masses presented e.g. in \cite{Rahimi:2022vlv, Moreno:2022goo}.
 It is worth emphasising that the coefficient of the Darwin operator is one order of magnitude larger than the corresponding ones at dimension-five.
However, as shown in detail in \cite{Lenz:2020oce}, this in fact follows from an accidental suppression of the dimension-five coefficients, rather than an enhancement of the Darwin term.
Therefore, the contribution of the Darwin operator,  neglected in previous phenomenological studies, turns out to be an important ingredient in the theoretical prediction of the $b$-baryon lifetimes, see section~\ref{sec:Num-analysis},
and of $B$ meson lifetimes \cite{Lenz:2022rbq}.
\begin{figure}[t]\centering
\includegraphics[scale=0.45]{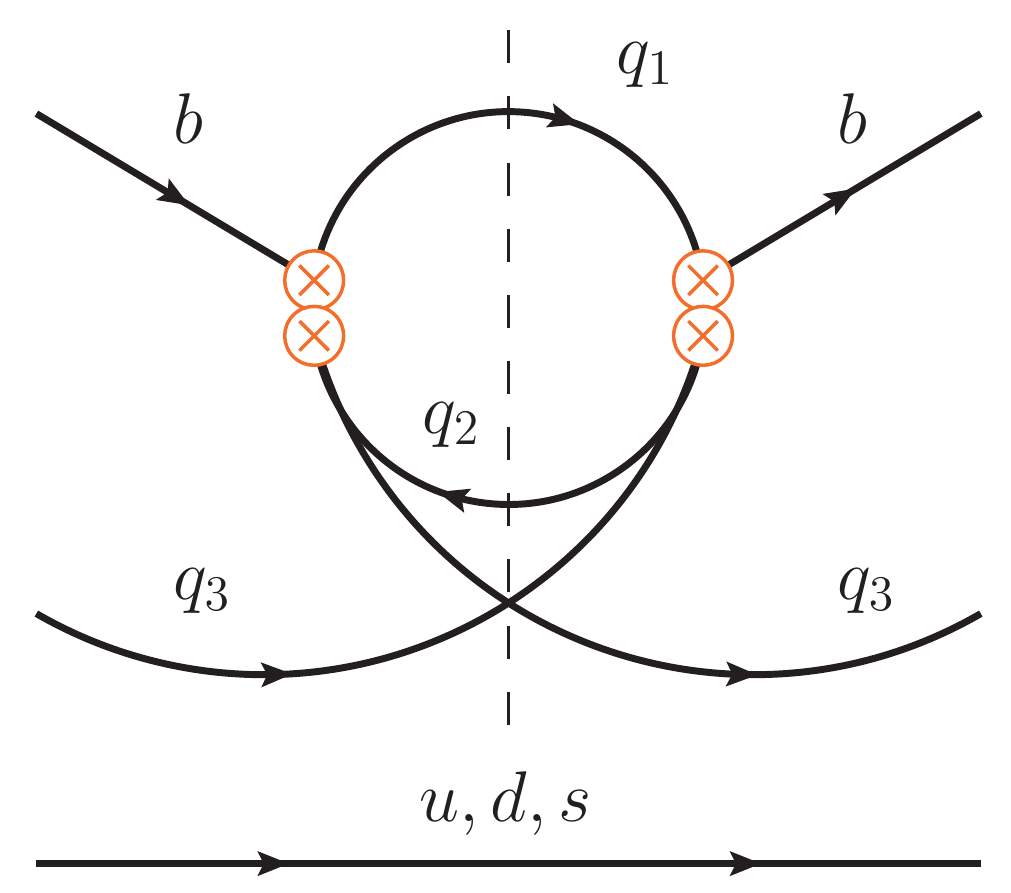}
\qquad \,
\includegraphics[scale=0.45]{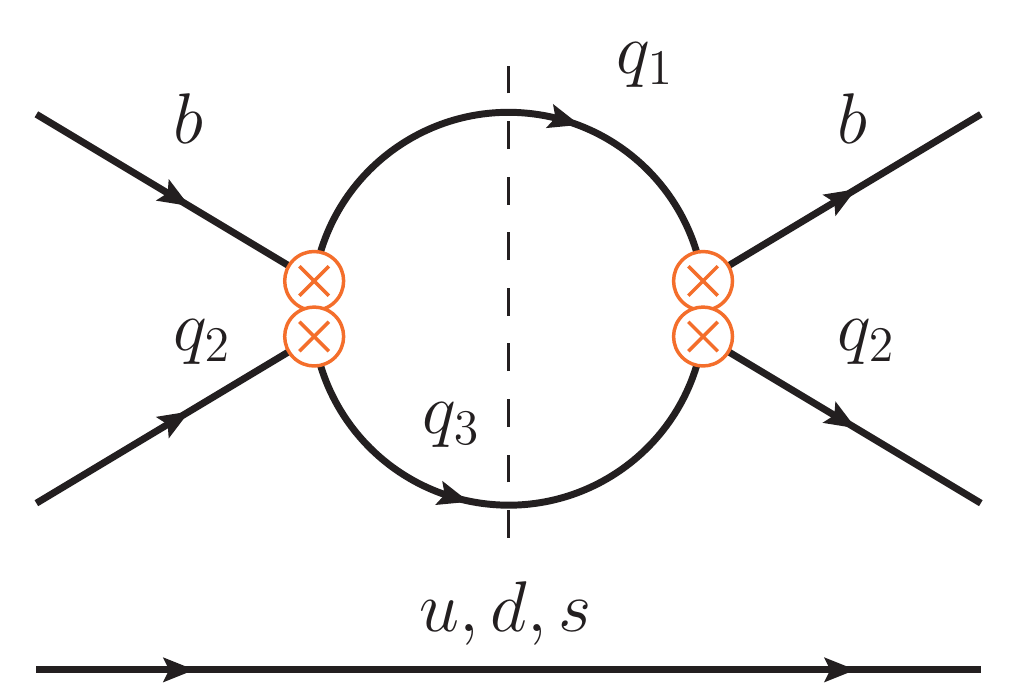}
\qquad \,
\includegraphics[scale=0.45]{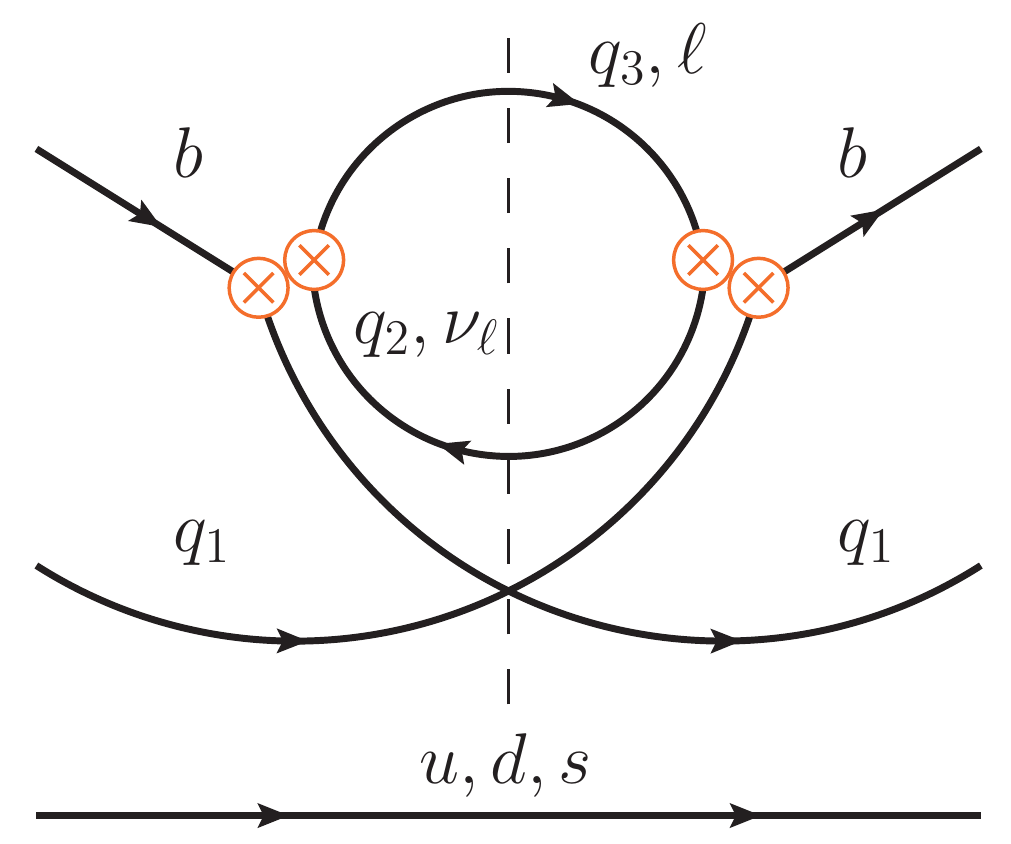}
\caption{Diagrams corresponding, from left to right, to the int$^-$, exc, and int$^+$ topologies at LO-QCD. 
The numbering scheme for quarks follows that in eq.~\eqref{eq:Q12}, so that $q_{1,2}$ are up-type quarks and $q_3$ is down-type. Note that semileptonic contributions only arise in the ${\rm int}^+$ topology. 
Disregarding the second, non-interacting spectator quark, these topologies are related by crossing to those in meson decays, cf. footnote \ref{ft:mes-bary-relation}. 
}
\label{fig:PI-WE-WA}
\end{figure}

The short-distance coefficients due to four-quark operators are obtained by computing, at LO-QCD, the discontinuity of the one-loop diagrams shown in figure~\ref{fig:PI-WE-WA}, commonly denoted in the literature as destructive Pauli interference (int$^-$), weak-exchange (exc), and constructive Pauli interference (int$^+$), respectively.\footnote{For $B$ mesons, the corresponding topologies are respectively denoted by weak-exchange (WE), Pauli interference (PI) and weak annihilation (WA). Hence, when translating results from baryons to mesons and vice-versa, the following replacements should be adopted: int$^- \leftrightarrow $ WE, exc $\leftrightarrow$ PI, and int$^+ \leftrightarrow$ WA \cite{Gratrex:2022xpm}. \label{ft:mes-bary-relation}}
Taking into account the different topologies, the dimension-six contribution from four-quark operators can be compactly written as
\begin{align}
16 \pi^2 \, \tilde{\Gamma}_6 \frac{\langle \tilde{\cal O}_6\rangle}{m_b^3}
 = \! \! \!
 \sum_{q_1,q_2,q_3}
\Biggl[
\tilde \Gamma_{6,{\rm int}^-}^{q_3} (x_{q_1}, x_{q_2}) +
\tilde \Gamma_{6,{\rm exc}}^{q_2} (x_{q_1}, x_{q_3}) +
\tilde \Gamma_{6,{\rm int}^+}^{q_1} (x_{q_2}, x_{q_3})
\Biggr]
\! + \!
\sum_{q_1, \ell}
\tilde \Gamma_{6,{\rm int}^+}^{q_1} (x_{\ell}, x_{\nu_\ell})
\,,
\label{eq:dim-6-4q-NLO-scheme}
\end{align}
where $\tilde \Gamma_{6,T}^q(x_{f_1},x_{f_2})$ denotes the imaginary part of the diagram
with topology
$T$, with external light quark~$q$ and internal fermions $f_1,f_2$, while $x_{f_i} = m_{f_i}^2/m_b^2$, where $m_{f_i}$ is the corresponding fermion mass. Note that
eq.~\eqref{eq:dim-6-4q-NLO-scheme} implies that, contrary to the corrections described so far, contributions to specific $b$-baryons differ not only due to different states appearing in the respective matrix elements, but also due to different short-distance coefficients.
In light of this, and of the formal loop enhancement with respect to two-quark operators, the effect of four-quark operators was expected to give the dominant correction to the total widths, and in particular to the lifetime ratios, see e.g. \cite{Uraltsev:1996ta,Neubert:1996we}. 
 
The functions $\tilde \Gamma_{6,T}^q(x_{f_1},x_{f_2})$ in eq.~\eqref{eq:dim-6-4q-NLO-scheme} can be further decomposed as follows:
\begin{align}
\tilde \Gamma_{6,T}^q(x_{f_1},x_{f_2})
 = \Gamma_0 
 \sum_{i = 1}^4
c^i_{6, T}(x_{f_1},x_{f_2}) \, \frac{\langle {{\cal O}}_i^{q} \rangle}{m_b^3} \,,
\label{eq:dim-6-4q-NLO}
\end{align}
with ${\cal O}^q_{1,\ldots, 4}$ denoting an appropriate set of four-quark operators, cf.\ eqs.~\eqref{eq:Dim6BaryonBasisHQE}-\eqref{eq:Dim6BaryonBasisHQE2}, and recall the short-hand notation $\langle  {\cal O}_i^q \rangle \equiv \langle {\cal B}|  {\cal O}_i^q | {\cal B}\rangle/(2 M_{\cal B})$. For a comprehensive discussion of these matrix elements, we refer to section~\ref{sec:matrixelements}.\footnote{Note that when comparing with the basis presented in section~\ref{sec:matrixelements}, one should make the identifications ${\cal O}^q_{3} = \tilde {\cal O}^q_{1}$ and ${\cal O}^q_{4} = \tilde {\cal O}^q_{2}$.}
The complete expressions for the coefficients $ c^{\,i}_{6,{\rm int}^-}(x_{q_1},x_{q_2})$ and $ c^{\,i}_{6, \rm exc}(x_{q_1},x_{q_3})$ up to NLO-QCD corrections, including also the contribution of 
the penguin and chromomagnetic operators, have been computed in \cite{Franco:2002fc} for four-quark operators defined in HQET.\footnote{Partial NLO results in the case of operators defined in QCD can be found in  \cite{Franco:2002fc, Beneke:2002rj}.}
The results for $ c^{\,i}_{6,{\rm int}^+}(x_{q_2},x_{q_3})$ can be
obtained, by means of a Fierz transformation,  
from the corresponding ones for 
$ c^{\,i}_{6, {\rm int}^-}(x_{q_1},x_{q_2})$ by replacing  $C_1 \leftrightarrow
C_2$, while for semileptonic modes, the NLO-corrections to the coefficients 
$ c^{\,i}_{6, {\rm int}^+}(x_{\ell}, x_{\nu_\ell})$ have been determined in \cite{Lenz:2013aua}. Because of the different terminology used to denote the same loop diagrams in baryons and mesons, in appendix~\ref{app:2} we present the LO-QCD expressions for the functions $\tilde \Gamma_{6,T}^q(x_{f_1},x_{f_2})$ given in eq.~\eqref{eq:dim-6-4q-NLO-scheme}.

Considering all possible contractions in the time-ordered product in eq.~\eqref{eq:optical_theorem}, the complete dimension-six four-quark operator contributions to $\Gamma(\cal B)$, included in our analysis, respectively read
\begin{align}
     16 \pi^2 \, \tilde{\Gamma}_6 \frac{\langle \tilde{\cal O}_6\rangle_{\Lab}}{m_b^3} & = 
      \Bigl[\tilde{\Gamma}^u_{6,\exc}(x_c, x_d)
     +  \tilde{\Gamma}^u_{6,\exc}(x_c,x_s)
     +  \tilde{\Gamma}^u_{6,\exc}(x_u,x_d)  
     +  \tilde{\Gamma}^u_{6,\exc}(x_u,x_s)
     \nonumber \\[2mm]
     &
     +  \tilde{\Gamma}^d_{6,\intm}(x_c,x_u)
     +  \tilde{\Gamma}^d_{6,\intm}(x_c,x_c)
     +  \tilde{\Gamma}^d_{6,\intm}(x_u,x_u)
     +  \tilde{\Gamma}^d_{6,\intm}(x_u,x_c)
     \nonumber \\[2mm]
     & 
     +  \tilde{\Gamma}^u_{6,\intp}(x_u,x_d)
     +  \tilde{\Gamma}^u_{6,\intp}(x_c,x_s)
     +  \tilde{\Gamma}^u_{6,\intp}(x_u,x_s) 
     +  \tilde{\Gamma}^u_{6,\intp}(x_c,x_d)
     \nonumber \\[2mm]
     &  
     +  \tilde{\Gamma}^{u}_{6,\intp}(x_\tau,x_{\nu_\tau})
     +  \tilde{\Gamma}^{u}_{6,\intp}(x_\mu,x_{\nu_\mu})
     +  \tilde{\Gamma}^{u}_{6,\intp}(x_e,x_{\nu_e})
     \Bigr]_{\Lab} \,,
    \label{eq:Gamma-Lb}
     \\[5mm]
    16 \pi^2 \, \tilde{\Gamma}_6 \frac{\langle \tilde{\cal O}_6\rangle_{\Xibo}}{m_b^3} & =
    \Bigl[
      \tilde{\Gamma}^u_{6,\exc}(x_c,x_d)
     +  \tilde{\Gamma}^u_{6,\exc}(x_c,x_s)
     +  \tilde{\Gamma}^u_{6,\exc}(x_u,x_d)
     +  \tilde{\Gamma}^u_{6,\exc}(x_u,x_s)
     \nonumber \\[2mm]
     &
     +  \tilde{\Gamma}^s_{6,\intm}(x_c,x_c)
     +  \tilde{\Gamma}^s_{6,\intm}(x_c,x_u)
     +  \tilde{\Gamma}^s_{6,\intm}(x_u,x_c)
     +  \tilde{\Gamma}^s_{6,\intm}(x_u,x_u)
     \nonumber \\[2mm]
     &
     +  \tilde{\Gamma}^u_{6,\intp}(x_u,x_d)
     +  \tilde{\Gamma}^u_{6,\intp}(x_c,x_s) 
     +  \tilde{\Gamma}^u_{6,\intp}(x_u,x_s)
     +  \tilde{\Gamma}^u_{6,\intp}(x_c,x_d)
     \nonumber \\[2mm] 
     & 
     +  \tilde{\Gamma}^{u}_{6,\intp}(x_\tau,x_{\nu_\tau})
     +  \tilde{\Gamma}^{u}_{6,\intp}(x_\mu,x_{\nu_\mu})
     +  \tilde{\Gamma}^{u}_{6,\intp}(x_e,x_{\nu_e})
     \Bigr]_{\Xibo} \,, 
    \label{eq:Gamma-Xi0}
    \\[5mm]
    16 \pi^2 \, \tilde{\Gamma}_6 \frac{\langle \tilde{\cal O}_6\rangle_{\Xibm}}{m_b^3} & =
    \Bigl[
     \tilde{\Gamma}^d_{6,\intm}(x_c,x_u) 
     +  \tilde{\Gamma}^s_{6,\intm}(x_c,x_c)
     + \tilde{\Gamma}^s_{6,\intm}(x_c,x_u)
     +  \tilde{\Gamma}^d_{6,\intm}(x_c,x_c)
     \nonumber \\[2mm]
     & 
     +  \tilde{\Gamma}^d_{6,\intm}(x_u,x_u)
     +  \tilde{\Gamma}^s_{6,\intm}(x_u,x_c) 
     +  \tilde{\Gamma}^s_{6,\intm}(x_u,x_u)
     +  \tilde{\Gamma}^d_{6,\intm}(x_u,x_c)
     \Bigr]_{\Xibm} \,, 
    \label{eq:Gamma-Xim}
    \\[5mm]
    16 \pi^2 \, \tilde{\Gamma}_6 \frac{\langle \tilde{\cal O}_6\rangle_{\Omb}}{m_b^3} & =
    \Bigl[
    \tilde{\Gamma}^s_{6,\intm}(x_c,x_c)
     +  \tilde{\Gamma}^s_{6,\intm}(x_c,x_u)
     +  \tilde{\Gamma}^s_{6,\intm}(x_u,x_c)
     +  \tilde{\Gamma}^s_{6,\intm}(x_u,x_u)
     \Bigr]_{\Omb} \,,
    \label{eq:Gamma-Om}
\end{align}
where we have now explicitly indicated the specific baryon appearing in the corresponding matrix elements. We stress that the results in eqs.~(\ref{eq:Gamma-Lb})-(\ref{eq:Gamma-Om}) do not take into account contributions in which the light quark in the four-quark operators differs from the spectator quarks in the $b$-baryon, the so-called `eye contractions'. These have been recently computed for mesons in \cite{King:2021jsq}, but they are still unknown for baryons. However, as they constitute subleading corrections to the dimension-six contribution, we expect their effect to go beyond the current accuracy of our study. Moreover, in our numerical analysis we only keep non-vanishing the masses of the charm quark and of the tau-lepton, i.e.\ we set $x_{u,d,s} = x_{\mu,e,\nu_{\ell}} = 0$, cf.\ eq.~\eqref{eq:xcandxtau}. 

Note that, in eqs.~(\ref{eq:Gamma-Lb})-(\ref{eq:Gamma-Om}), the non-leptonic contributions have been ordered by topology, and within each topology we have listed the terms in order of their CKM hierarchy. In particular, the leading contributions to the $\Lab$ and $\Xibo$ decay widths arise from the $\intm$ and $\exc$ topologies. 
As for the semileptonic contributions, they can only arise in the int$^+$ topology, and, since we do not include the eye contractions, they only enter the decay width of the $\Lab$ and $\Xibo$ baryons, see the last line of eqs.~(\ref{eq:Gamma-Lb}), (\ref{eq:Gamma-Xi0}), and not that of the $\Xibm$ or $\Omb$. 
However, the semileptonic contributions give a negligible numerical effect to the total widths, and in particular  do not generate any significant splitting between the semileptonic branching fractions of $b$-baryons, as expected because of the strong CKM suppression $|V_{ub}|^2 \ll |V_{cb}|^2$. Thus, within our current sensitivity, any difference between the semileptonic branching fractions of $b$-baryons can arise only from SU(3)$_F$ effects in the matrix elements of the two-quark operators.

In section~\ref{sec:Num-analysis}, we present our predictions for the lifetimes ratios of the $b$-baryons with the $B_d$ meson. For completeness, in order to facilitate the comparison, the corresponding leading dimension-six four-quark contribution for the latter is \cite{Lenz:2022rbq} 
\begin{align}
    16 \pi^2 \, \tilde{\Gamma}_6 \frac{\langle \tilde{\cal O}_6\rangle_{B_d}}{m_b^3} & =
    \Bigl[
     \tilde{\Gamma}^d_{6,{\rm WE}}(x_c,x_u)
     +  \tilde{\Gamma}^d_{6,{\rm WE}}(x_c,x_c)
     +  \tilde{\Gamma}^d_{6,{\rm WE}}(x_u,x_u)
     +  \tilde{\Gamma}^d_{6,{\rm WE}}(x_u,x_c)
     \Bigr]_{B_d}.
    \label{eq:Gamma-Bd}
\end{align}

Finally, at order $1/m_b^4$, the short-distance contributions due to four-quark operators are also known in the literature, albeit only at LO-QCD, see e.g.\ \cite{King:2021xqp}. They have been determined in~\cite{Gabbiani:2003pq, Gabbiani:2004tp} 
for operators defined in QCD\footnote{Some inconsistencies in the expressions of \cite{Gabbiani:2003pq,Gabbiani:2004tp} were identified in \cite{Lenz:2022rbq}, cf.\ footnote 8 therein.} 
and also in~\cite{Lenz:2013aua, Lenz:2022rbq} for the HQET operators. 
However, as compared with our previous studies \cite{King:2021xqp,Gratrex:2022xpm,Lenz:2022rbq}, we do not include the subleading $1/m_b$ corrections to the four-quark matrix elements in our central values for the total widths, preferring instead to treat these contributions as part of the uncertainty estimate. 
The reason for this is the absence of a consistent procedure to determine the corresponding matrix elements for baryons, particularly in HQET, due to a proliferation of the dimension-seven operators, see e.g.~\cite{Neubert:1993mb}, and in contrast to the case of mesons, where the vacuum insertion approximation~(VIA) provides 
a first estimate. This problem was extensively discussed in~\cite{Gratrex:2022xpm}.
Moreover, other $1/m_b^4$ corrections are also missing, namely those due to two-quark operators, which so far are known only for semileptonic $b$-quark decays~\cite{Dassinger:2006md,Mannel:2010wj,Gambino:2016jkc,Mannel:2018mqv}, and those to the
dimension-six matrix elements, see section~\ref{sec:matrixelements}. As a result, a complete analysis of the $b$-baryon lifetimes up to this order is currently not possible. 
 Given that, in the $b$-system, power corrections prove to be well under control, we consider it more justified in this work to treat the $1/m_b^4$ contributions as an additional source of uncertainty, rather than trying to include them in the central values for lifetimes
with only partial, and potentially misleading, estimates for the dimension-seven matrix elements.

\subsection{Non-perturbative Matrix Elements}
\label{sec:matrixelements}

In this section, we present our determinations of the hadronic parameters. It is convenient to first consider the matrix elements of the four-quark operators, followed by the discussion of the two-quark matrix elements $\mupi{\Bary},\,\muG{\Bary},\, \rhoD{\Bary}$.

A basis of dimension-six four-quark operators in HQET suitable for the $b$-baryons is \cite{Neubert:1996we}\footnote{The recent study \cite{Gratrex:2022xpm} made use of the QCD basis of operators instead.}
\begin{alignat}{3}
\Opsixprime{1}{q} &= (\bar{h}_{v}^i \gamma_\mu (1- \gamma_5) q^i) (\bar{q}^j \gamma^\mu (1 - \gamma_5) h_{v}^j)\,, \qquad
& \Opsixprime{2}{q} &= (\bar{h}_{v}^i (1-\gamma_5) q^i) (\bar{q}^j (1+\gamma_5) h_{v}^j) \,, 
\label{eq:Dim6BaryonBasisHQE}
\\[1mm]
\Opsixtprime{1}{q} &= (\bar{h}_{v}^i \gamma_\mu(1-\gamma_5)   q^j) (\bar{q}^j \gamma^\mu(1-\gamma_5)   h_{v}^i)\,, \qquad
&\Opsixtprime{2}{q} &= (\bar{h}_{v}^i(1-\gamma_5) q^j) (\bar{q}^j (1+\gamma_5) h_{v}^i) \,, \label{eq:Dim6BaryonBasisHQE2}
\end{alignat} 
with $q$ labeling the light quark in the corresponding operator, i.e.\ $q = u,d,s $.
Note that the colour-rearranged operators $\tilde {\cal O}_{1,2}^q$ 
are related to the colour-octet ones commonly adopted in studies of heavy meson lifetimes, see e.g.\ \cite{King:2021xqp, Gratrex:2022xpm, Lenz:2022rbq}, by the completeness property of the SU(3)$_c$ generators 
\begin{equation}
    t_{ij}^a t^a_{lm} = \frac12 \Bigl( \delta_{im}\delta_{jl} - \frac{1}{N_c} \delta_{ij}\delta_{lm} \Bigr)\,.
    \label{eq:coloridentity}
\end{equation}
The usefulness of the choice of basis in eqs.~\eqref{eq:Dim6BaryonBasisHQE}, \eqref{eq:Dim6BaryonBasisHQE2} is exhibited by the relations  
\begin{equation}
\langle\mathcal{B}|\Opsixtprime{i}{q}|\mathcal{B}\rangle = - \tilde{B}_i^q \langle\mathcal{B}| \Opsixprime{i}{q} |\mathcal{B}\rangle \,, \qquad i = 1,2\,,
\label{Eq:BaryonMERelation1}
\end{equation}
where, assuming the valence quark approximation, the total colour antisymmetry of the baryon wave function imposes $\tilde{B}_i^q=1$ \cite{Neubert:1996we}. In our study, we consider a universal parameter $\tilde B_i^q \equiv \tilde B$,\footnote{In general, $\tilde B_1^q = \tilde B_2^q + {\cal O} (1/m_b)$, and $\tilde B^{u,d} \neq \tilde B^s$, but we neglect these subleading corrections.} with $\tilde{B}=1$ valid at a typical hadronic scale $\mu_h \ll m_b$.
When performing the numerical analysis, we vary this scale in the range $1 \, {\rm GeV} \leq \mu_h \leq 1.5 \, {\rm GeV}$, while taking as our reference value  $\mu_h = 1.5\, $ GeV. 

In order to estimate the matrix elements on the r.h.s.\ of eq.~\eqref{Eq:BaryonMERelation1}, 
we adopt the non-relativistic constituent quark model (NRCQM), according to which the matrix elements of the colour-singlet four-quark operators can be expressed, in terms of baryon wave functions evaluated at the origin, as
\begin{align}
    \cfrac{\langle {\cal T}_b| \Opsixprime{1}{q}|{\cal T}_b\rangle}{2M_{{\cal T}_b}} &= -|\Psi^{{\cal T}_b}(0)|^2 \, , \quad\quad\quad \cfrac{\langle {\cal T}_b| \Opsixprime{2}{q}|{\cal T}_b\rangle}{2M_{{\cal T}_b}} = 
    \frac{1}{2}|\Psi^{{\cal T}_b}(0)|^2\,, \label{Eq:4q1}
\end{align}
for the SU(3)$_F$ triplet ${\cal T}_b =(\Lab, \Xibm, \Xibo)$, and
\begin{align}
    \cfrac{\langle \Omb| \Opsixprime{1}{s}|\Omb\rangle}{2M_{\Omb}} &= -6|\Psi^{\Omb}(0)|^2 \, , \quad\quad\quad \cfrac{\langle \Omb| \Opsixprime{2}{s}|\Omb\rangle}{2M_{\Omb}} = -|\Psi^{\Omb}(0)|^2\,,
    \label{Eq:4q2}
\end{align}
for the $\Omb$. It should be emphasised that the constituent quark picture provides access only to the valence quark contributions, for which the field of a light quark within the operator matches at least one of the baryon valence quarks. The missing non-valence contributions are, however, expected to provide subleading corrections. Hence, in \eqref{Eq:4q1}, it should be understood that the relations are valid only when the light quark $q$ in the operator $\Opsixprime{i}{q}$ matches one of the valence quarks in the baryon ${\cal T}_b$, and the matrix element is otherwise taken to be zero, and similarly in \eqref{Eq:4q2} for the $\Omb$.

We stress that,
apart from the exploratory study in \cite{DiPierro:1999tb}, which has never been followed up, there are no lattice determinations  for the four-quark baryonic matrix elements available. A computation for the $\Lab$, within HQET sum rules, 
was performed in \cite{Colangelo:1996ta}. In contrast to the case of $B$ mesons, where one can set up a sum rule for the small deviation of the bag parameter from one \cite{King:2021jsq,King:2019lal,Kirk:2017juj,Grozin:2016uqy},  for baryons one can only write down sum rules for the whole matrix element. Thus, the baryon case
may be sensitive to stability issues often associated with three-point sum rules \cite{Braun:1999dp}.
Moreover, the sum rule work in \cite{Colangelo:1996ta}
does not yet include NLO-QCD effects.
These corrections can be large, as was shown in the HQET sum rule calculation of the two-point correlator~\cite{Yakovlev:1996bs}, entering also the computation of the four-quark matrix element, where the $\alpha_s$-contributions appear to be of a similar size to the leading contribution. Very recently, the four-quark $\Lab$ matrix elements were also determined with QCD sum rules \cite{Zhao:2021lzd}, confirming the relatively small values obtained by the HQET sum rules in \cite{Colangelo:1996ta}.\footnote{A separate sum rule calculation in \cite{Huang:1999xj} was able to accommodate the then very low experimental values of the $\tau(\Lab)/\tau(B_d)$ lifetime ratio, at the expense of an anomalously large four-quark contribution; see figure~\ref{fig:tau-Lambda-b-to-tau-Bd-history-Exp-vs-Th}.}
We are not aware of sum rule determinations of the matrix elements of the $\Xibo$,  $\Xibm$, or $\Omb$ baryons. 
Therefore, in this work, we choose to consistently apply the NRCQM 
to calculate the matrix elements of the dimension-six four-quark operators for all the baryons considered. For comparison, however, we briefly discuss the numerical impact of the sum rule determination from \cite{Colangelo:1996ta} on the $\Lab$ lifetime in section~\ref{sec:Num-analysis}.

Following the standard approach proposed by de Rujula, Georgi, and Glashow \cite{DeRujula:1975qlm}, 
the baryon wave functions can be extracted from the known values of hyperfine mass splittings~\cite{Barger:1979pd,Cortes:1980jz, Blok:1991st,Rosner:1996fy}. 
In the NRCQM, the hyperfine splittings are controlled by the short-distance gluon exchange between the constituent quarks. For a generic hadron $H$, the mass $M_H$ can be expressed as
\begin{equation}
    M_H=M_0+\langle H_{\text{spin}}\rangle\,,
    \label{Eq:MH}
\end{equation}
where $M_0$ contains the spin-independent contributions, including the constituent quark masses and the binding energies. The spin-dependent terms are, for the ground state $(L=0)$ hadrons, given as 
 \begin{align}
 H_{\rm spin, \, baryons} &= \sum_{i > j} \frac{16 \pi\alpha_s }{9} \frac{(\vec{s}_i\cdot\vec{s}_j)}{m_i^{\, \bary} \, m_j^{\, \bary}} \delta^3(\vec{r}_{ij})  \,, 
 \label{eq:HFmass-bar}\\
 H_{\rm spin, \, mesons} &= \frac{32 \pi\alpha_s}{9}  \frac{(\vec{s}_i\cdot\vec{s}_j)}{m_i^{\, \mes} \, m_j^{\, \mes}} \delta^3(\vec{r}_{ij})  \,,
 \label{eq:HFmass1}
 \end{align}
 where $i,j,$ label the constituent quarks in the hadron, with  masses $m^{\, \bary}_i$ and $m^{\, \mes}_i$ respectively for baryons and mesons, while $\vec{s}_i$ denotes the corresponding quark spin operator. 
 When evaluating the expectation value in eq.~\eqref{Eq:MH} for a given hadronic state, the delta functions in eqs.~\eqref{eq:HFmass-bar}, \eqref{eq:HFmass1} result in the modulus squared of the hadron wave function at the origin, $|\Psi^H(0)|^2$, and the light quarks in $b$-baryons are taken to form a diquark spin state. 
 Note that we do not assume the constituent quark masses within mesons and baryons to be equal, i.e.\  $m_i^{\, \mes} \neq m_i^{\, \bary}$, but instead take their values as used in the fit to hadronic masses \cite{Karliner:2014gca}. 
 
  Following the approach of Rosner \cite{Rosner:1996fy}, 
  the wave functions appearing in eqs.~\eqref{Eq:4q1}, \eqref{Eq:4q2} are extracted using the hyperfine splittings between the positive-parity spin-3/2 and spin-1/2 bottom baryons. For example, for the $\Lambda_b$ baryon, this results in the relation\footnote{Applying a similar relation for mesons, and using this to estimate the decay constant, would lead to the estimates $f_B = 0.188(14) \GeV,\, f_{B_s} = 0.241(18) \GeV$, where the uncertainty arises from varying the scale of $\alpha_s(\mu_h)$ between 1.0 and 1.5 GeV. These values are consistent with those obtained from lattice computations, supporting the applicability of the NRCQM to baryons.}
 \begin{equation}
     M_{\Sigma_b^\ast}- M_{\Sigma_b}=\frac{16\pi\alpha_s}{m_b^{\,\bary}\, m_{\tilde q}^{\, \bary}} \, \frac{3}{2}|\Psi^{{\Lambda_b}}(0)|^2\,,
 \end{equation}
 with $\tilde q = u,d,$ while the corresponding relations for $|\Psi^{\Xi_b}(0)|^2$, and $|\Psi^{\Omega_b}(0)|^2$, involve the hyperfine splittings $M_{\Xi_b^\ast}- M_{\Xi'_b}$, and $M_{\Omega_b^\ast}- M_{\Omega_b}$, respectively. After normalising these relations to the analogous expressions involving the meson mass splittings, we can express the matrix elements in eqs.~\eqref{Eq:4q1}, \eqref{Eq:4q2}, in terms of $B$-meson wave functions\footnote{Note that we use interchangeably the notation $B_u = B_d \equiv B$, in the limit of exact isospin symmetry.} as  
\begin{align}
\displaystyle
    \frac{\langle \Lambda_b\vert \Opsixprime{1}{ \tilde q}\vert \Lambda_b\rangle}{2 M_{\Lambda_b}}& = -y_{\tilde q} \, \frac{4}{3}\frac{M_{\Sigma_b^\ast}- M_{\Sigma_b}}{M_{B^\ast}-M_{B}}|\Psi^{{B}}(0)|^2\,,
    \label{Eq:4qMatrixElements1}\\[0.8em]
   \frac{\langle \Xibo\vert \Opsixprime{1}{u}\vert \Xibo\rangle}{2 M_{\Xi_b}} &= \frac{\langle \Xibm\vert  \Opsixprime{1}{d}\vert \Xibm \rangle}{2 M_{\Xi_b}}=-y_{\tilde q} \, \frac{4}{3}\frac{M_{\Xi_b^\ast}- M_{\Xi'_b}}{M_{B^\ast}-M_{B}}|\Psi^{{B}}(0)|^2\,, \label{Eq:4qMatrixElements2}\\[0.8em]
   \frac{\langle \Xibm\vert  \Opsixprime{1}{s}\vert \Xibm \rangle}{2 M_{\Xi_b}}&=\frac{\langle \Xibo\vert  \Opsixprime{1}{s}\vert \Xibo \rangle}{2 M_{\Xi_b}}=-y_s \, \frac{4}{3}\frac{M_{\Xi_b^\ast}- M_{\Xi'_b}}{M_{B_s^\ast}-M_{B_s}}|\Psi^{B_s}(0)|^2\,, \label{Eq:4qMatrixElements3}\\[0.8em]
   \frac{\langle \Omb\vert \Opsixprime{1}{s}\vert \Omb\rangle}{2 M_{\Omega_b}} &= -y_s 
   \, 6 \, \frac{4}{3}
   \frac{M_{\Omega_b^\ast}- M_{\Omega_b}}{M_{B_s^\ast}-M_{B_s}}|\Psi^{B_s}(0)|^2\,,
   \label{Eq:4qMatrixElements4}
\end{align}
where $y_{\tilde q}$, $y_s$, denote ratios of the constituent quark masses in baryons and mesons~\cite{Karliner:2014gca}
\begin{equation}
y_{\tilde q} = \frac{m_b^{\, \bary} \, m_{\tilde q}^{\, \bary}}{m_b^{\, \mes} \, m_{\tilde q}^{\, \mes}} \simeq 1.18\,, \qquad \qquad
y_s = \frac{m_b^{\, \bary} \, m_s^{\, \bary}}{m_b^{\, \mes} \, m_s^{\, \mes}} \simeq 1.12 \,.
\label{Eq:ConstituentMassRatio}
\end{equation}
The ratios of the mass splittings,
\begin{equation}
    r_{q}(\Bary)\equiv \frac{4}{3}\frac{M_{\Bary^\ast}-M_\Bary}{M_{B_{q}^\ast}-M_{B_{q}}}\,, 
\end{equation}
are key inputs for the evaluation of the matrix elements.\footnote{In \cite{Jenkins:1996de,Jenkins:1996rr}, both $1/m_Q$ and $1/N_c$ expansions were employed simultaneously, leading to the exact relation $r_q(\Bary) \to 2/3$ in the $m_Q, N_c \to \infty$ limit. This holds quite accurately in the charm sector, and could also be expected to apply, in principle to an even greater degree of accuracy, in the $b$ sector. However, the relationship is potentially sensitive to higher-order corrections in the $1/N_c$ expansion.} In our numerical analysis we use the experimental values of meson and baryon mass splittings, when available~\cite{Workman:2022ynf}, and assume exact isospin symmetry within the hyperfine splittings, i.e. $M_{B^{\ast}_d}-M_{B_d}=M_{B^{\ast}_u}-M_{B_u}$. As for the ratio $r_s(\Omb)$, since the mass of the $\Omega_b^\ast$ has not yet been measured, we employ the result for the splitting  $M_{\Omega^\ast_b}-M_{\Omega_b}$ 
from~\cite{Karliner:2014gca}, consistently with the use of the values of the constituent mass ratios in eq.~\eqref{Eq:ConstituentMassRatio}. This leads to
\begin{equation}
    r_s(\Omega_b)=0.66\pm 0.22\,.\label{Eq:rsOmega}
\end{equation}
A comparison between the predictions for $r_{q}$, based both on NRCQM fits and lattice QCD evaluations, alongside the corresponding available experimental results, is shown in table \ref{Tab:rq}.
We note that for the $B$-meson mass splittings, we use the averages of the experimental values reported in~\cite{Workman:2022ynf}.
\begin{table}[H]
\renewcommand{\arraystretch}{1.3}
\centering
\begin{tabular}{|c||c|c||c|}
\hline
Quantity & Experiments \cite{Workman:2022ynf}& Lattice QCD \cite{Brown:2014ena}  & NRCQM~\cite{Karliner:2014gca}  \\ \hline
$r_{\tilde q}(\Sigma_b)$ & $ 0.58 \pm 0.01$ & $ 0.62\pm 0.26$ & $0.63 \pm 0.24$ \\
\hline
$r_{\tilde q}(\Xi'_b)$& $ 0.60 \pm 0.00$ & $ 0.79\pm 0.27$ & $0.67 \pm 0.24$ \\
\hline
$r_s(\Xi'_b)$ & $0.56\pm 0.02$& $0.74\pm 0.25$  & $0.63\pm 0.22$ \\
\hline
$r_s(\Omega_b)$ & unknown  & $0.78\pm 0.22$ & $0.66\pm 0.22$\\
\hline
\end{tabular}
\caption{\small Comparisons of the NRCQM results for the $r_{q}(\Bary)$ to available experimental data and lattice QCD evaluations. For the $B$-meson mass splittings, we use the measured values reported in~\cite{Workman:2022ynf}.} 
\label{Tab:rq}
\end{table}

Having the ratios of hadron mass splittings under control, we proceed by relating the meson wave functions in eqs.~(\ref{Eq:4qMatrixElements1})-(\ref{Eq:4qMatrixElements4}) to the static decay constants via
\begin{equation}
    |\Psi^{B_{q}}(0)|^2=\frac{F_{B_{q}}^2(\mu_0)}{12}\,, 
    \label{eq:wave-fuction-decay-const}
\end{equation}
with
\begin{equation}
    \langle 0\vert\bar{q}\gamma^\mu\gamma_5 h_v\vert B_{q}\rangle_{\text{HQET}}=i\,F_{B_q}(\mu_0)\sqrt{M_{B_{q}}} \, v^\mu\,,
\end{equation}
following the conventions for the HQET states used in~\cite{King:2021xqp}.
Assuming the constituent-quark relations for the matrix elements of the operators $\mathcal{O}^q_i$ in eqs.~\eqref{Eq:4q1}, \eqref{Eq:4q2}, as well as the valence quark approximation result $\tilde{B}=1$ in \eqref{Eq:BaryonMERelation1}, to be satisfied at a low hadronic scale $\mu_h$, in eq.~\eqref{eq:wave-fuction-decay-const} we set $\mu_0 = \mu_h=1.5\,\text{GeV}$, the same hadronic scale that was used in the HQET sum rule derivation of the corresponding bag parameters in $B$ mesons~\cite{King:2021jsq}. 
The value of the static decay constant at the scale $\mu_h$ can be extracted using its relation~\cite{Neubert:1992fk} to the QCD decay constant in the static limit $\hat{f}_{B_{q}}$, 
\begin{equation}
    \hat{f}_{B_{q}}=\frac{F_{B_{q}}(\mu_0)}{\sqrt{M^{\phantom{-}}_{B_{q}}}}\bigg[1+\frac{\alpha_s(\mu_0)}{2\pi}\bigg(\ln\frac{\mu_b^2}{\mu_0^2}-\frac{4}{3}\bigg)\bigg]\,, \label{eq:staticfqalphas}
\end{equation}
where $\hat{f}_{B_{q}}$ differs from the full QCD decay constant $f_{B_{q}}$ used for meson lifetimes,  by the terms of order $\mathcal{O}(1/m_b)$, and $\mu_b=4.5\,\text{GeV}$.
The parameter $\hat{f}_{B_{q}}$ is available from lattice QCD simulations~\cite{Aoki:2014nga}, from which we take the numerical values  
\begin{equation}
    \hat{f}_{B}= (219\pm 17)\,\text{MeV}\,,\qquad \qquad \hat{f}_{B_s}= (264\pm 19)\,\text{MeV}\,,
\end{equation}
which result in
\begin{equation}
\begin{split}
    F_{B}(\mu_h=1.5\,\text{GeV})=(0.48\pm 0.04)\,\text{GeV}^{3/2}\,,\\
    F_{B_s}(\mu_h=1.5\,\text{GeV})=(0.58\pm 0.04)\,\text{GeV}^{3/2}\,,
\end{split}
\end{equation}
as compared to $F_{B}(\mu_b)=(0.53\pm 0.04)\,\text{GeV}^{3/2}$ and $F_{B_s}(\mu_b)=(0.64\pm 0.05)\,\text{GeV}^{3/2}$.
With this ingredient in place, we list in table~\ref{Tab:O1Numerics} the numerical values of the relevant matrix elements of the operator ${\cal O}_1^q$ at the scale $\mu_h$.
\begin{table}[ht]
\renewcommand{\arraystretch}{1.7}
\centering
\begin{tabular}{|c||c|}
\hline
Matrix elements at $\mu_h=1.5\,\text{GeV}$& Value in units $\text{GeV}^3$\\
\hline
$\displaystyle{\langle \Opsixprime{1}{u}\rangle_{\Lab} = \langle \Opsixprime{1}{d}\rangle_{\Lab}}$ & 
$ -0.013 \pm 0.002 \pm { 0.004}$ \\[0.0em]
$\displaystyle{\langle \Opsixprime{1}{u}\rangle_{\Xibo}} = \langle \Opsixprime{1}{d}\rangle_{\Xibm}$ & 
$ -0.014\pm 0.002 \pm { 0.004}$ \\
$\displaystyle{\langle \Opsixprime{1}{s}\rangle_{\Xibo} = \langle \Opsixprime{1}{s}\rangle_{\Xibm}}$ & 
$-0.018\pm 0.003 \pm { 0.005} $\\
\hline
$\displaystyle{\langle \Opsixprime{1}{s}\rangle_{\Omb}}$ & $-0.126\pm 0.046 \pm { 0.038}$\\
\hline
\end{tabular}
\caption{\small
Numerical values for the matrix elements of the operator $\Opsixprime{1}{q}$ at the scale $\mu_h=1.5\,\text{GeV}$, following the notation $\langle {\cal O}_i^q \rangle_{\mathcal{B}} \equiv \langle {\cal B}| {\cal O}_i^q | {\cal B}\rangle/(2 M_{\cal B})$.  The first errors are obtained by varying the input parameters, and the second ones by adding a conservative $30\%$ model uncertainty. The remaining matrix elements are, at the scale $\mu_h$, related to the matrix elements of $\mathcal{O}^q_1$ via eqs.~\eqref{Eq:BaryonMERelation1}, \eqref{Eq:4q1}, and \eqref{Eq:4q2}, with $\tilde{B}_i^q(\mu_h)=1$. } 
\label{Tab:O1Numerics}
\end{table}
Using the results for the renormalisation group evolution of the matrix elements of the dimension-six four-quark operators within HQET \cite{Shifman:1984wx,Voloshin:1986dir,Politzer:1988wp,Neubert:1996we}, for $\mu_h=1.5\,\GeV$ and $\mu_b=4.5\,\GeV$, we obtain  
\begin{equation}
\begin{pmatrix}
\langle {\cal O}_1^q \rangle \\
\langle {\cal O}_2^q \rangle \\
\langle \tilde {\cal O}_1^q \rangle \\
\langle \tilde {\cal O}_2^q\rangle
\end{pmatrix}\!(\mu_b)=
\begin{pmatrix}
1.29 && 0 && -0.09 && 0 \\
0 && 1.29 && 0 && -0.09 \\
0 && 0 && 1 && 0 \\
0 && 0 && 0 && 1
\end{pmatrix}\begin{pmatrix}
\langle {\cal O}_1^q \rangle \\
\langle {\cal O}_2^q \rangle\\
\langle \tilde {\cal O}_1^q \rangle \\
\langle \tilde {\cal O}_2^q \rangle
\end{pmatrix}\!(\mu_h)\,.
\end{equation}
Then, at the scale $\mu_b$, the matrix elements for the triplet ${\cal T}_b$ and the $\Omega_b$ baryon read respectively
\begin{equation}
\begin{split}
    \begin{pmatrix}
\langle \Opsixprime{1}{q}\rangle \\
\langle \Opsixprime{2}{q}\rangle \\
\langle \Opsixtprime{1}{q}\rangle \\
\langle \Opsixtprime{2}{q}\rangle
\end{pmatrix}_{\mathcal{T}_b}\!\!\!\!(\mu_b)=\begin{pmatrix}
1.38 \, \langle  \Opsixprime{1}{q}\rangle \\
-0.69 \, \langle \Opsixprime{1}{q}\rangle \\
-\langle \Opsixprime{1}{q}\rangle \\
\frac{1}{2}\langle \Opsixprime{1}{q}\rangle
\end{pmatrix}_{\mathcal{T}_b}\!\!\!\!(\mu_h)\,,\qquad\qquad
\begin{pmatrix}
\langle \Opsixprime{1}{s}\rangle \\
\langle \Opsixprime{2}{s}\rangle \\
\langle \Opsixtprime{1}{s}\rangle \\
\langle \Opsixtprime{2}{s}\rangle
\end{pmatrix}_{\Omega_b}\!\!\!\!\!(\mu_b)=\begin{pmatrix}
1.38 \, \langle  \Opsixprime{1}{s}\rangle \\
0.23 \, \langle \Opsixprime{1}{s}\rangle \\
-\langle \Opsixprime{1}{s}\rangle \\
-\frac{1}{6}\langle \Opsixprime{1}{s}\rangle
\end{pmatrix}_{\Omega_b}\!\!\!\!\!(\mu_h)\,,
\end{split}
\label{eq:MEs-mh-to-mb}
\end{equation}
which amounts, for both the triplet and the $\Omega_b$, to a modification of the parameter $\tilde{B}$ from the value $\tilde{B}(\mu_h)=1$ to $\tilde{B}(\mu_b)=1.38$.\footnote{Choosing the value for the initial scale $\mu_h=1\,\text{GeV}$ instead of $\mu_h=1.5\,\text{GeV}$ results in $\tilde{B}(\mu_b)=1.66$.} At the same time, the one-loop running preserves the ratios between the matrix elements of the operators $\Opsixprime{1}{q}$ and $\Opsixprime{2}{q}$.

We now turn to discuss the remaining, non-spectator matrix elements \cite{Bigi:1993ex,Bigi:1994ga,Mannel:1994kv,Dassinger:2006md},
\begin{align}
\mupi{\Bary} &= - \frac{1}{2M_\Bary} \langle \Bary |\bar{b}_v (i D_\mu) (i D^\mu) b_v | \Bary \rangle \,,
\label{eq:mu-PI}
\\
\muG{\Bary} &=\frac{1}{2M_\Bary} \langle \Bary |\bar{b}_v (i D_\mu) (i D_\nu) (- i \sigma^{\mu \nu}) b_v  |\Bary \rangle\,,
\label{eq:mu-G}
\\
\rhoD{\Bary} &= \frac{1}{2M_\Bary}\langle \Bary |\bar{b}_v (i D_\mu) (i v \cdot D) (i D^\mu) b_v | \Bary \rangle \,,
\label{eq:rho-D}
\end{align} 
which correspond to the kinetic, chromomagnetic, and Darwin parameters respectively. Following~\cite{Dassinger:2006md}, we define the operators in terms of the field $b_v(x)$, rather than the HQET field $h_v(x)$, with differences due to this choice arising only at order $1/m_b^4$. These parameters can be further related to the heavy-quark expansion of the hadron mass \cite{Falk:1992wt,Falk:1992ws,Bigi:1994ga,Neubert:1996qg},
\begin{equation}
M_\Bary = m_b + \bar{\Lambda} + \frac{\mupi{\Bary}}{2m_b} - \frac{\muG{\Bary}}{2m_b} + \mathcal{O}\left(\frac{1}{m_b^2}\right) \,,
\label{eq:mHexpand}
\end{equation}
where $\bar{\Lambda} \sim 0.5 \GeV$. 
Applying the expansion \eqref{eq:mHexpand} to the mass difference between hyperfine partners, and taking into account the proportionality of the chromomagnetic parameter to the spin factor $d_{\cal B}$, we have
\begin{equation}
\label{eq:la2def2}
\muG{\Bary} = d_\Bary \frac{M_{\Bary^{*}}^2 - M_\Bary^2}{d_\Bary - d_{\Bary^{*}}} \,,
\end{equation}
with
\begin{equation}
\label{eq:dHdef}
d_\Bary = -2 \left(S_\Bary (S_\Bary+1) - S_b (S_b +1 ) - S_l(S_l+1)\right) \,,
\end{equation}
and $S_X$ denoting the spin of the particle $X$.
As only $d_{\Omb}$ is non-zero, with $d_{\Omb}=4$ and $d_{{\Omb}^*}=-2$, it follows that $\muG{\Bary}=0$ for the triplet ${\cal T}_b$, while, using the masses and splitting from \cite{Karliner:2014gca}, we obtain
\begin{equation}
    \muG{\Omb} = (0.193 \pm 0.065 \pm 0.019) \GeV^2 \,.
\end{equation}
Here, the first uncertainty is parametric, while the second one corresponds to our $10\%$~uncertainty estimate from missing higher-order $1/m_b$ corrections.

Concerning the kinetic parameter, one can relate $\mupi{\Lab}$ to $\mupi{B}$ via 
\begin{equation}
    \overline{M}_B-M_{\Lab}=\bar{\Lambda}_B-\bar{\Lambda}_{\Lab}+\frac{\mupi{B}-\mupi{\Lab}}{2m_b}+\mathcal{O}\left(\frac{1}{m_b}\right) \,,
\end{equation}
where we differentiate between the parameter $\bar{\Lambda}$ for mesons and baryons, and $\overline{M}_B$ denotes the spin-averaged meson mass $\overline{M}_B=(M_B+3 M_{B^\ast})/4$. To proceed, we assume the equality of the difference $\bar{\Lambda}_{B_q}-\bar{\Lambda}_\Bary$ in the bottom and charmed sectors, as well as $\mupi{B}= \mupi{D}$ and $\mupi{\Lab}=\mupi{\Lac}$, resulting in the expression~\cite{Bigi:1992su,Bigi:1995jr}
\begin{equation}
    \left(\overline{M}_D  -  M_{\Lac} \right) - \left(\overline{M}_B  -  M_{\Lab} \right) =  \left( \frac{1}{2m_c} - \frac{1}{2m_b} \right) \left( \mupi{B} - \mupi{\Lab} \right)\\ +  \mathcal{O}\left(\frac{1}{m_{b}},\frac{1}{m_{c}}\right) \,,\label{Eq:mupiLambda}
\end{equation}
where $\overline{M}_D = (M_D + 3 M_{D^\ast})/4$, and for the inputs on the left-hand side, we have used the isospin-averaged hadron masses. Unlike in the charm sector, however, there have been analyses of inclusive semileptonic $B \to X_c \, \ell \nu_\ell$ decays~\cite{Alberti:2014yda,Gambino:2016jkc,Bordone:2021oof,Bernlochner:2022ucr} in order to extract the values of the parameter $\mupi{B}$ from fits to experimental data. For our numerical analysis, we use the value obtained in~\cite{Bordone:2021oof}:
\begin{equation}
    \mupi{B}=\left( 0.477\pm 0.056\right) \GeV^2\,.
    \label{Eq:mupiB}
\end{equation}
Furthermore, we adopt the spectroscopic estimate of the size of SU(3)$_F$-breaking from~\cite{Bordone:2022qez, Lenz:2022rbq}:
\begin{equation}
    \mupi{B_s}-\mupi{B}= \left( 0.04\pm 0.02\right) \GeV^2\,.
    \label{Eq:mupimu}
\end{equation}

\begin{table}[t]
\renewcommand{\arraystretch}{1.4}
    \centering
    \begin{tabular}{|c||c|c|c|} \hline
        & $\Lab$ & $\Xi_b^{0,-}$  & $\Omb$  \\ \hline
      $\muG{\Bary}/ \GeV^2$  
      & 0 
      & 0 
      & $0.193 \pm 0.068$ 
      \\ 
      $\mupi{\Bary}/ \GeV^2$ 
      & $0.50 \pm 0.06$ 
      & $0.54 \pm 0.06$ 
      & $0.56 \pm 0.06$ 
      \\
      $\rhoD{\Bary}/ \GeV^3$ 
      & $0.031 \pm 0.009$  
      & $0.037 \pm 0.009$ 
      & $0.050 \pm 0.021$ \\ \hline
    \end{tabular}
    \caption{Non-perturbative parameters for the non-spectator contributions used in our analysis. Values for $\mupi{\Bary}$ follow from the relations derived in eqs.~\eqref{Eq:mupiLambda} and \eqref{Eq:mupiOmega}, with $\mupi{B}$ taken from the fit value in \cite{Bordone:2021oof} and $\mupi{B_s}$ obtained using the SU(3)$_F$-breaking estimate from \cite{Bigi:2011gf}. Values for $\rhoD{\Bary}$ follow from employing the equation of motion~\eqref{eq:DarwinEoM}. The errors quoted here are obtained by combining in quadrature the parametric uncertainty and the uncertainty due to missing power corrections.}
\label{Tab:parametrs2}
\end{table}
\noindent
For the $\Omega_b$, the analogous relation, derived for the first time in \cite{Gratrex:2022xpm}, is
\begin{equation}
\label{eq:mupiOmega2}
    \mupi{\Omb}\left(\frac{1}{2m_b} - \frac{1}{2m_c} \right) \simeq m_c - m_b + \frac{1}{3}\left( \left(M_{\Omb} + 2 M_{{\Omb}^{*}} \right) - \left(M_{\Omc} + 2 M_{{\Omc}^{*}} \right)\right) 
    +  \mathcal{O}\left(\frac{1}{m_{b}},\frac{1}{m_{c}}\right)\,.
\end{equation}
This can be recast, using a similar relation for the $B_s$ meson, as
\begin{equation}
    \left(\mupi{\Omb}-\mupi{B_s}\right)\left(\frac{1}{2m_b} - \frac{1}{2m_c} \right)  \simeq\overline{M}_{D_s} - \overline{M}_{B_s} + \frac{1}{3}\left( \left(M_{\Omb} + 2 M_{{\Omb}^{*}} \right) - \left(M_{\Omc} + 2 M_{{\Omc}^{*}} \right)\right) \,, \label{Eq:mupiOmega}
\end{equation}
up to corrections of order $1/m_{b,c}$, and 
where $\overline{M}_{B_s}$, $\overline{M}_{D_s}$ are again spin-averaged meson masses. Concerning the quark masses, we use their values in the kinetic scheme, i.e.\ $m_b^{\rm kin}(\mu_{\text{cut}}=1\,\GeV)=4.57\,\GeV$, as extracted from the fit in~\cite{Bordone:2021oof,Bordone:2022qez}, and $m^{\rm kin}_c(\mu_{\text{cut}}=0.5\GeV)=1.40\,\GeV$ \cite{Fael:2020iea,Chetyrkin:2000yt,Herren:2017osy}.  
The values for the differences of the kinetic parameters turn out to be small, and we obtain\footnote{This small separation between $\mupi{B}$ and $\mupi{\Lab}$ is consistent with the sum rules calculation in \cite{Colangelo:1995qp}.}
\begin{align}
    \mupi{\Lab} - \mupi{B} & = (0.029 \pm 0.001 \pm 0.015)\GeV^2 \,,  \\
    \mupi{\Xi_b} - \mupi{B} & = (0.061 \pm 0.002 \pm 0.030) \GeV^2 \,,  \\
    \mupi{\Omb} - \mupi{B_s} & = \left( 0.040 \pm 0.023 \pm 0.020\right) \GeV^2 \,, 
\end{align}
where again the first quoted errors represent the parametric uncertainties, while the second ones follow from our assignment of $50\%$ uncertainties to account for possibly sizeable $1/m_c$ corrections. Combining the above results with those in \eqref{Eq:mupiB}, \eqref{Eq:mupimu} leads to our estimates for the baryonic kinetic parameters presented in table~\ref{Tab:parametrs2}. 

As for the Darwin parameter
$\rhoD{\Bary}$, this can be related, up to $\mathcal{O}(1/m_b)$ corrections, to the four-quark matrix elements by the equation of motion for the gluon field strength tensor,
\begin{equation}
 \left[i D_\mu, i D_\nu \right]= i g_s G_{\mu \nu} \,, \qquad   [D^\mu, G_{\mu \nu}] = - g_s t^a \sum_{q=u,d,s}\bar{q} \gamma_\nu t^a q \,, 
\end{equation}
which leads to the relation 
\begin{equation}
    2 M_\Bary \, \rhoD{\Bary}  = 
    g_s^2 
    \sum_{q=u,d,s}\langle \Bary | \left( -\frac{1}{8}\Opsixprime{1}{q} + \frac{1}{24}\Opsixtprime{1}{q} +  \frac{1}{4}\Opsixprime{2}{q} - \frac{1}{12}\Opsixtprime{2}{q}  \right) | \Bary \rangle + \mathcal{O} \left( \frac{1}{m_b} \right) \,,
    \label{eq:DarwinEoM}
\end{equation}
in terms of the operator basis defined in \eqref{eq:Dim6BaryonBasisHQE}. 
We evaluate the right-hand side of eq.~\eqref{eq:DarwinEoM} using the matrix elements of the four-quark operators renormalised at the scale $\mu_0=\mu_b$, which, together with $\alpha_s(\mu_b) = 0.22$, results in the values for the Darwin parameter shown in table~\ref{Tab:parametrs2}.\footnote{For comparison, the values obtained using instead  $\alpha_s=1$, and the four-quark matrix elements at the low hadronic scale $\mu_h$, read $(\rhoD{\Lambda_b},\rhoD{\Xi_b}, \rhoD{\Omega_b})\simeq (0.11, 0.13, 0.18)\,\GeV^3$. However, the limit on the size of $\rhoD{B}$ derived in \cite{Chow:1995mz} supports a lower value of $\alpha_s$, consistent with the results in table~\ref{Tab:parametrs2}.}
\section{Numerical Analysis and Results}
\label{sec:Num-analysis}

\begin{figure}[ht]
    \centering
    \includegraphics[scale=1.4]{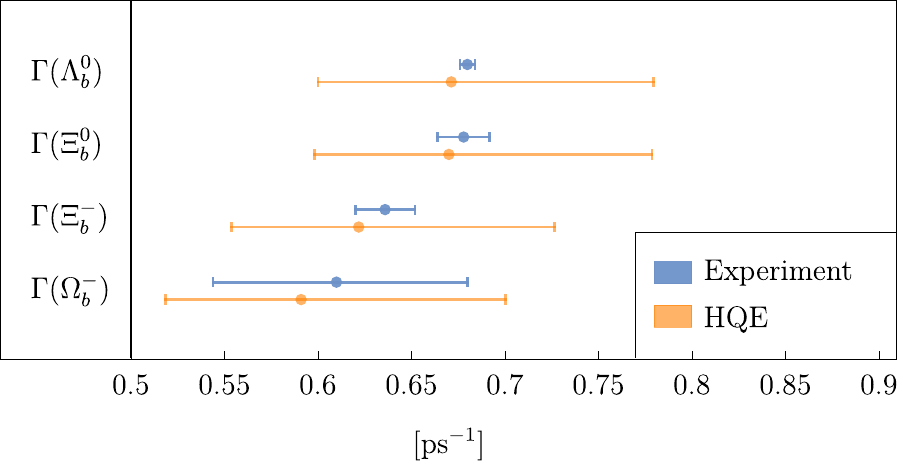} 
    \includegraphics[scale=1.4]{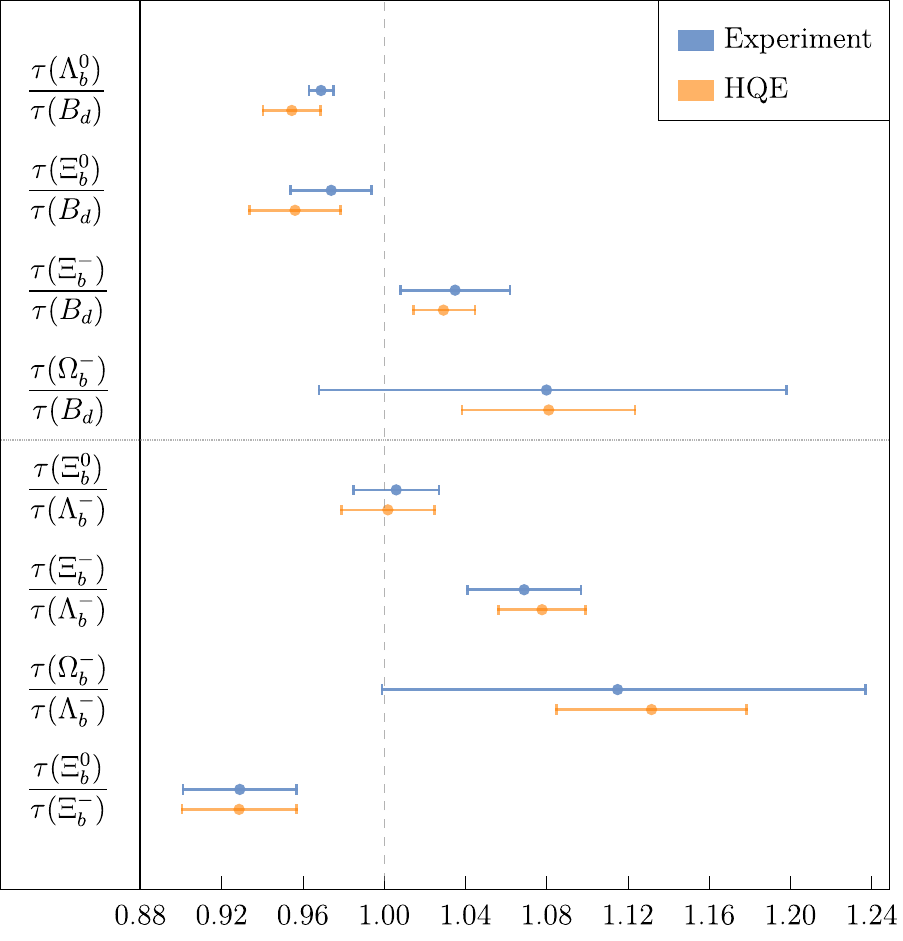}
    \caption{Graphical representation of the results presented in table~\ref{tab:HQE-vs-Data}.}
    \label{fig:HQE-vs-Data}
\end{figure}

In this section, we present our predictions for the total decay
widths of $b$-baryons and their lifetime ratios, as well as for the values of their lifetimes normalised to $\tau(B_d)$, as summarised in table~\ref{tab:HQE-vs-Data} and figure~\ref{fig:HQE-vs-Data}. We also provide results for the semileptonic decay widths and inclusive $b$-baryon semileptonic branching fractions, shown in eqs.~\eqref{eq:Gamma-SL} and \eqref{eq:SL-BF-1}, \eqref{eq:SL-BF-2}.

The values of the non-perturbative parameters used in our numerical analysis are displayed in tables~\ref{Tab:O1Numerics} and~\ref{Tab:parametrs2} of section~\ref{sec:matrixelements}, while all remaining inputs are collected in 
appendix~\ref{app:1}. Note that the renormalisation scales $\mu_1$ and $\mu_0$ are varied independently, both in the same interval $\mu_b/2 \le \mu_{0,1} \le 2 \mu_b$, with $\mu_b = 4.5\GeV$, and using as central values $\mu_{0} =\mu_1 = \mu_b$.
In addition, in order to account for possible uncertainties in our assumption for the ``factorisation'' scale~$\mu_h$, we vary this 
between $1\GeV$ and $1.5\GeV$, fixing its central value to  $1.5 \GeV$. 

As we present results for the lifetime ratios of $b$-baryons with the $B_d$ meson, a couple of comments with respect to our recent study \cite{Lenz:2022rbq} are in order. Firstly,
as discussed in section~\ref{sec:short-dist-contr},
in our analysis of the $b$-baryon total widths, we treat the dimension-seven contributions as an additional source of uncertainty, and do not provide any estimates for their central values. Hence, for consistency, here we have adopted the same treatment also for the total width of the $B_d$ meson, differently from~\cite{Lenz:2022rbq}.\footnote{However, for the $B_d$ meson, the dimension-seven four-quark contribution turns out to be negligible, see eq.~(3.4) of~\cite{Lenz:2022rbq}.} Secondly, as the value of the Darwin parameter $\rho_D^3({\cal B})$ for baryons is obtained using the equations of motion for the gluon field strength tensor evaluated at the scale $\mu_b$, see eq.~\eqref{eq:DarwinEoM}, we again follow the same procedure for the $B_d$ meson and use\footnote{This value is consistent with the experimental fit in \cite{Bernlochner:2022ucr}, rather than that in \cite{Bordone:2021oof}.}
\begin{equation}
\rho_D^3 (B_d) = (0.028 \pm 0.010) \, {\rm GeV}^3.     
\end{equation}
 Our predictions for the total widths are determined from eq.~\eqref{eq:HQE}, while the lifetime ratios are obtained using the relation
\begin{equation}
\frac{\tau(H_1)}{\tau(H_2)} 
= 1 + \left[\Gamma(H_2) - \Gamma(H_1)\right]^{\rm HQE} \tau(H_1)^{\rm exp}\,,
\label{eq:defRatio}
\end{equation}
where the difference $\Gamma(H_2) - \Gamma(H_1)$ is computed from eq.~\eqref{eq:HQE}, and 
we use as input the experimental value for the lifetime of the $H_1$ hadron. 

In order to understand the size of each of the contributions in the HQE included in our analysis, 
below we show our results for the decomposition of the total widths of $b$-baryons, explicitly indicating the LO- and NLO-QCD corrections when the latter are present. For central values of the input parameters, we obtain 
\begin{align}
\Gamma (\Lab) & =  
\Gamma_0 \Biggr[ (\, \underbrace{5.97}_{\rm LO} - \underbrace{0.44}_{\rm \Delta NLO})
- 0.14 \, \frac{\mu_\pi^2 (\Lab)}{\GeV^2} 
- 1.35 \, \frac{\rho_D^3 (\Lab)}{\GeV^3} 
- \big(\underbrace{10.6}_{\rm LO} + \underbrace{5.04}_{\rm \Delta NLO} \! \big)
\frac{\langle \Opsixprime{1}{q}\rangle_{\Lab}}{\GeV^3}
\Biggl] \,,  
 \label{eq:Lambda-res} 
\\[3mm]
\Gamma (\Xibo) & =  
\Gamma_0 \Biggr[(\, \underbrace{5.97}_{\rm LO} - \underbrace{0.44}_{\rm \Delta NLO} )
- 0.14 \, \frac{\mu_\pi^2 (\Xibo)}{\GeV^2}
- 1.35 \, \frac{\rho_D^3 (\Xibo)}{\GeV^3} 
\nonumber \\
&  
\qquad\qquad \qquad - \big(\underbrace{18.2}_{\rm LO} + \underbrace{4.02}_{\rm \Delta NLO}\! \big)
\frac{\langle \Opsixprime{1}{q}\rangle_{\Xibo}}{\GeV^3}
- \big(\underbrace{-7.31}_{\rm LO} + \underbrace{1.48}_{\rm \Delta NLO}\! \big)
\frac{\langle \Opsixprime{1}{s}\rangle_{\Xibo}}{\GeV^3}
\Biggl]\,, 
  \\[3mm]
\Gamma (\Xibm) & =  
\Gamma_0 \Biggr[(\, \underbrace{5.97}_{\rm LO} - \underbrace{0.44}_{\rm \Delta NLO} )
- 0.14 \, \frac{\mu_\pi^2 (\Xibm)}{\GeV^2}
- 1.35 \, \frac{\rho_D^3 (\Xibm)}{\GeV^3} 
\nonumber \\
&  
\qquad\qquad \qquad - \big(\underbrace{-7.62}_{\rm LO} + \underbrace{1.02}_{\rm \Delta NLO}\!\big)
\frac{\langle \Opsixprime{1}{q}\rangle_{\Xibm}}{\GeV^3}
- \big(\underbrace{-7.31}_{\rm LO} + \underbrace{1.48}_{\rm \Delta NLO}\!\big)
\frac{\langle \Opsixprime{1}{s}\rangle_{\Xibm}}{\GeV^3}
\Biggl] \,,
\end{align}
\begin{align}
\Gamma (\Omb) & =  
\Gamma_0 \Biggr[ (\, \underbrace{5.97}_{\rm LO} - \underbrace{0.44}_{\rm \Delta NLO} )
- 0.14 \, \frac{\mu_\pi^2 (\Omb)}{\GeV^2} 
- 0.24 \, \frac{\mu_G^2 (\Omb)}{\GeV^2}
- 1.35 \, \frac{\rho_D^3 (\Omb)}{\GeV^3} 
\nonumber 
\\
&
\qquad \qquad \qquad 
- \big(\underbrace{-3.81}_{\rm LO} + \underbrace{0.72}_{\rm \Delta NLO}\!\big)  \frac{\langle \Opsixprime{1}{s}\rangle_{\Omb}}{\GeV^3}
\Biggl]\,,
\end{align}
with $q = u, d$.
The total decay widths are clearly dominated by the dimension-three contribution, with the radiative corrections giving a ${\sim} 10$\% effect. 
Among the power-suppressed terms, the largest contribution comes from dimension-six four-quark operators, and in particular from the exc topology, which enters the $\Lab$ and $\Xibo$ widths. Radiative corrections also play an important role, and range from ${\sim} 10$\% to ${\sim} 50$\% of the four-quark contribution depending on the specific topology. The Darwin term gives the next dominant power correction, and in some cases partially compensates the contribution of four-quark operators, as for example in the $\Lab$, eq.~\eqref{eq:Lambda-res}.

For completeness, we also show the decomposition for the total width of the $B_d$ meson, cf.\ eq.~(3.4) of~\cite{Lenz:2022rbq}: 
	\begin{eqnarray}
		\Gamma (B_d^0) & = &
		\Gamma_0
		\biggl[ 
		(\, \underbrace{5.97}_{\rm LO} - 
		\underbrace{0.44}_{\rm \Delta  NLO} )
		- \, 0.14 \, \frac{\mu_{\pi}^2 (B)}{\rm GeV^2}
		- 0.24 \, \frac{\mu_{G}^2 (B)}{\rm GeV^2}
		- 1.35 \, \frac{\rho_{D}^3 (B)}{\rm GeV^3}
		\nonumber
		\\[2mm]
		& &
		\quad 
		- \, (\, \underbrace{0.012}_{\rm LO} + \underbrace{0.022}_{\rm \Delta NLO}\,) \,  {\tilde B}_1^{q} 
		+ (\,\underbrace{0.012}_{\rm LO} + \underbrace{0.020}_{\rm \Delta NLO}\,) \,  {\tilde B}_2^{q}
		- \, (\,\underbrace{0.74}_{\rm LO} + \underbrace{0.03}_{\rm \Delta NLO}) \,  {\tilde B}_3^{q} 
		\nonumber \\[3mm]
		& & \quad + \, (\,\underbrace{0.78}_{\rm LO} - \underbrace{0.01}_{\rm \Delta NLO}) \,  {\tilde B}_4^{q} 
		- 0.14 \,  \tilde \delta^{q q^\prime}_{1} 
		+ 0.02 \,  \tilde \delta^{q q^\prime}_{2} 
		- 2.29 \,  \tilde \delta^{q q^\prime}_{3} 
		+ 0.00 \,  \tilde \delta^{q q^\prime}_{4} 
		\nonumber
		\\[1mm]
		& &  
		\quad
		- \, 0.01 \,  \tilde \delta^{sq}_{1} 
		+ 0.01 \,  \tilde \delta^{sq}_{2} 
		- 0.69 \,  \tilde \delta^{sq}_{3} 
		+ 0.78 \,  \tilde \delta^{sq}_{4}\,
		\biggr] \,, 
		\label{eq:Gammad}
	\end{eqnarray}
where $\tilde B_i^{q}$ and $\tilde \delta_i^{q q^\prime}, \tilde \delta_i^{s q}$
denote, respectively, the $B$ meson dimension-six Bag parameters and the
`eye contractions', see \cite{Lenz:2022rbq} for details. Their numerical values, as well as of those for $\mu_\pi^2 (B)$ and $\mu_G^2 (B)$,  are taken to be the same as in \cite{Lenz:2022rbq}. 

Our HQE predictions for the $b$-baryon lifetimes and their ratios, together with the corresponding experimental values, are presented in table~\ref{tab:HQE-vs-Data} and visualised in figure~\ref{fig:HQE-vs-Data}. The quoted theoretical errors are obtained by combining uncertainties due to variation of the input parameters and of the renormalisation scales $\mu_0, \mu_1$, and $\mu_h$, as well as an additional $15\%$ uncertainty added to the dimension-six contribution to account for missing $1/m_b^4$ corrections.
Overall, we find excellent agreement between the HQE predictions and the experimental data for all the observables considered. 

It is important to point out that computing the lifetime ratios entirely within the HQE, i.e.\ without using the experimental values for $\tau(H_1)^{\rm exp} $ in eq.~\eqref{eq:defRatio}, 
leads to very similar results as those in table~\ref{tab:HQE-vs-Data}, 
albeit with slightly larger uncertainties.
Furthermore, when using the HQET sum rules result for the four-quark matrix elements \cite{Colangelo:1996ta}
\begin{equation}
\displaystyle{\langle \Opsixprime{1}{u}\rangle_{\Lambda_b} = \langle \Opsixprime{1}{d}\rangle_{\Lambda_b}} = - (3.2 \pm 1.6) \times 10^{-3} {\rm \GeV}^3 \,, 
\end{equation}
we obtain a larger value for the lifetime ratio $\tau (\Lab)/ \tau (B_d^0)$, namely
\begin{equation}
\tau (\Lab)/ \tau (B_d^0)  = 0.976 \pm 0.012 \,,
\end{equation}
which however is consistent, within uncertainties, with the value shown in table~\ref{tab:HQE-vs-Data}. 

Finally, we also present HQE predictions for the inclusive semileptonic decay rates $\Gamma_{\rm SL} ({\cal B})$, defined~as
\begin{equation}
\Gamma_{\rm SL} ({\cal B}) \equiv 
\Gamma ({\cal B} \to X_{c+u} \ell \bar \nu_\ell)\,, 
\end{equation}
with a massless lepton $\ell = e, \mu$. 
We obtain
\begin{align}
    \Gamma_{\rm SL}({\cal T}_b ) &= 0.075^{+0.004}_{-0.003} \,\, {\rm ps}^{-1}\,,
    \qquad\qquad  \Gamma_{\rm SL}(\Omega_b ) =0.073^{+0.004}_{-0.003} \, \,{\rm ps}^{-1}\,,
\label{eq:Gamma-SL}
\end{align}
which leads to the following results for the inclusive semileptonic branching fractions ${\rm BR}_{\rm SL} ({\cal B})$:
\begin{align}
{\rm BR}_{\rm SL}(\Lab) & =  
    (11.0^{+0.6}_{-0.5}) \, \% \,,
    \qquad\qquad 
    {\rm BR}_{\rm SL} (\Xibm) =  
    (11.7^{+0.7}_{-0.6}) \, \%,
    \label{eq:SL-BF-1} \\[2mm]
    {\rm BR}_{\rm SL} (\Xibo) & =  
    (11.1^{+0.6}_{-0.6}) \, \%  \,,\qquad\qquad
    {\rm BR}_{\rm SL}(\Omb) =  
    (12.0^{+1.4}_{-1.4}) \, \% \,,
    \label{eq:SL-BF-2}
\end{align}
where 
\begin{equation}
{\rm BR}_{\rm SL} ({\cal B}) = \Gamma_{\rm SL} ({\cal B}) \, \tau({\cal B})^{\rm exp} \,.
\end{equation}
Note that the value for ${\rm BR}_{\rm SL} (\Lab) $ in eq.~\eqref{eq:SL-BF-1} perfectly agrees with the result obtained in the recent study~\cite{Colangelo:2020vhu}. Although measurements of inclusive $b$-baryon semileptonic branching fractions are extremely difficult at present machines, the theoretical predictions might still prove useful in Monte Carlo simulations. 

\begin{table}[th]
\renewcommand{\arraystretch}{1.5}
\centering
\begin{tabular}{|c||c|c|}
    \hline
    Observable & \quad HQE prediction \quad & \quad Experimental value \quad \\
\hhline{|=||=|=|}
    $\Gamma (\Lab)$ &  
    $0.671^{+0.108}_{-0.071} \, \psin$ &
    $( 0.680 \pm 0.004) \, \psin $
    \\
     \hline
    $\Gamma (\Xibo)$ &  
    $0.670^{+0.108}_{-0.071} \, \psin$ &
    $( 0.678 \pm 0.014) \, \psin $
    \\
    \hline
    $\Gamma (\Xibm)$ &  
    $0.622^{+0.104}_{-0.067} \, \psin$ &
    $(0.636 \pm 0.016) \, \psin $
    \\
    \hline
    $\Gamma (\Omb)$ &  
    $0.591^{+0.108}_{-0.071} \, \psin$ &
    $0.610^{+0.070}_{-0.066} \, \psin $
    \\
\hhline{|=||=|=|}
    $\tau (\Lab)/ \tau (B_d^0)$  &  
    $0.955 \pm 0.014$ &
    $0.969 \pm 0.006$ 
    \\
    \hline
    $\tau (\Xibo)/ \tau (B_d^0)$ &  
    $0.956 \pm 0.023$ &
    $0.974 \pm 0.020^{\, *}$ 
    \\
    \hline
    $\tau (\Xibm)/ \tau (B_d^0)$ &  
    $1.029 \pm 0.015$ &
    $1.035 \pm 0.027^{\, *}$ 
    \\
    \hline
    $\tau (\Omb)/ \tau (B_d^0)$ &  
    $1.081 \pm 0.042$ &
    $1.080^{+0.118 \, *}_{-0.112}$  
    \\
\hhline{|=||=|=|}
    $\tau (\Xibo)/ \tau (\Lab)$ &  
    $1.002 \pm 0.023$ &
    $1.006 \pm 0.021^{\, *}$ 
    \\
    \hline
     $\tau (\Xibm)/ \tau (\Lab)$ &  
    $1.078 \pm 0.021$ &
    $1.069 \pm 0.028^{\, *}$ 
    \\
    \hline
    $\tau (\Omb)/ \tau (\Lab)$ &  
    $1.132 \pm 0.047$ &
    $1.115^{+0.122 \, *}_{-0.116}$ 
    \\
    \hline
    $\tau (\Xibo)/ \tau (\Xibm)$ &  
    $0.929 \pm 0.028$ &
    $0.929 \pm 0.028$ 
    \\
    \hline
\end{tabular}
\caption{
Comparison between our predictions based on the HQE and the data. The theoretical uncertainties are obtained by combining uncertainties due to input parameters, the renormalisation scales $\mu_0, \mu_1$, and $\mu_h$, and missing $1/m_b^4$ corrections.  
The experimental numbers marked with an asterisk are obtained by dividing the corresponding values shown in table~\ref{tab:exp-data}, and do not take into account possible experimental correlations.  
}
\label{tab:HQE-vs-Data}
\end{table}

\section{Conclusions}
\label{sec:conclusions}
We have performed a phenomenological study of the lifetimes of $b$-baryons, including for the first time the contribution of the Darwin operator and a new extraction of the matrix elements of the four-quark operators within the framework of the non-relativistic constituent quark model.
Overall, we observe an excellent agreement between our predictions and the experimental data. For the lifetime ratios $\tau(\Lab)/ \tau (B_d)$,
$\tau(\Xibo)/ \tau (B_d)$, and $\tau(\Xibo)/ \tau (\Xibm)$, the theoretical and experimental uncertainties are comparable, 
while for the total decay rates the theoretical errors dominate, although, in the case of the $\Omb$ baryon, the experimental uncertainties are also still quite sizeable.

In particular, we find 
\begin{eqnarray}
\frac{\tau(\Lab)}{\tau (B_d^0)}
^{\rm HQE} = 1 - (0.045 \pm 0.014)
\, ,
& \quad &
\frac{\tau(\Lab)}{\tau (B_d^0)}
^{\rm Exp.} = 1 - (0.031 \pm 0.006)
\, ,
\end{eqnarray}
showing that the measured suppression of the $\Lab$ lifetime by $(-3.1 \pm 0.6)\%$ compared to $\tau(B_d)$ is impressively confirmed by the corresponding theory prediction of $(-4.5 \pm 1.4)\%$.
Therefore, we do not see any indications for visible violations of quark-hadron duality affecting the HQE, as applied to the $\Lab$ baryon.

It is interesting to note that the theory estimate from 1986 \cite{Shifman:1986mx} led to almost exactly the same central value as the one obtained in our study. 
The authors of \cite{Shifman:1986mx} included in their analysis: LO-QCD corrections to the free-quark decay, $\Gamma_3^{(0)}$ in eq.~(\ref{eq:HQE}), taking into account charm quark mass dependence; LO-QCD corrections to the spectator effects, $\tilde \Gamma_6^{(0)}$ in eq.~(\ref{eq:HQE}), without charm quark mass dependence; and estimates of the matrix elements of the four-quark operators based on a simplified version of the non-relativistic constituent quark model. They neglected corrections of order $1/m_b^2$, i.e.\ $ \Gamma_5^{(0)}$ in eq.~(\ref{eq:HQE}), as well as $1/N_c$ corrections in the free quark decay. Furthermore, the NLO-QCD corrections to the $\Delta B= 1$ Wilson coefficients, to the free-quark decay, $\Gamma_3^{(1)}$ in 
eq.~\eqref{eq:HQE}, and to the spectator effects, $\tilde \Gamma_6^{(1)}$ in eq.~\eqref{eq:HQE}, as well as the contribution of the Darwin operator, $ \Gamma_6^{(0)}$ in eq.~\eqref{eq:HQE},  
were unknown in 1986. Shifman and Voloshin correctly predicted 36 years ago a small negative deviation of $\tau (\Lab) / \tau (B_d^0)$ from one, however, the perfect matching of their result with our post-diction from 2023 is a kind of numerical coincidence, since the effect of their approximations seems to have cancelled with the low value of the decay constant used in 1986, $f_{B} = 110$ MeV, resulting in $\left(f_{B}^{(1986)}/f_{B}^{(2023)}\right)^2 \approx 0.34$.

Moreover, we confirm the experimentally observed lifetime splitting of the
 $\Xibo$ and $\Xibm$ baryons
\begin{eqnarray}
\frac{\tau(\Xibo)}{\tau (\Xibm)}
^{\rm HQE} = 1 - (0.071 \pm 0.028)
\, ,
& \quad &
\frac{\tau(\Xibo)}{\tau (\Xibm)}
^{\rm Exp.} = 1 - (0.071 \pm 0.028)
\, ,
\end{eqnarray}
coincidentally obtaining the same central value and uncertainty estimate. For the $\Omb$ baryon we predict a larger lifetime compared to the $B^0_d$ meson, although here a clear experimental confirmation is still missing.

Our results also agree, within uncertainties, with the most recent estimate of $b$-baryon lifetimes presented in \cite{Cheng:2018rkz}. This agreement holds in spite of the fact that NLO-corrections in dimension-six four-quark contributions, as well as the Darwin contribution, which was at the time unknown, are missing from the theoretical expression in \cite{Cheng:2018rkz}, while the uncertainties in \cite{Cheng:2018rkz} are artificially small, as they arise only from the variation of $\mu_h$ and from the bag parameters entering the four-quark contribution to $\tau(B_d^0)$, and do not include other parametric and scale uncertainties.

Concerning the lifetime hierarchy, our calculations indicate
\begin{eqnarray}
\tau (\Lab)
\approx
\tau (\Xibo)
<
\tau (\Xibm)
\leq
\tau (\Omb) \, ,
\end{eqnarray}
which is confirmed by data. Note that this hierarchy was already predicted in e.g. \cite{Guberina:1999bw,Cheng:2018rkz}.

Finally, we have presented numerical updates for the inclusive semileptonic branching fractions of the $b$-baryons, which currently seem to be difficult to measure at LHCb, and are
not possible at $\Upsilon (4S)$ runs with Belle II. However, they might be feasible for the flavour physics programme at the high luminosity upgrade of the LHC \cite{Cerri:2018ypt}
or even further in the future at FCC-ee, see e.g. 
\cite{Lenz:2021bkv, Lenz:2022kal}.

In order to further improve the theoretical precision in the lifetime ratios, the following calculations can be performed in the future:
\begin{itemize}
\item[$*$] Non-perturbative, and in particular lattice QCD, determinations of the matrix elements of the four-quark operators of dimension-six, $\langle \tilde{ \mathcal{O}}_6 \rangle $, and of
dimension-seven, $\langle \tilde{ \mathcal{O}}_7 \rangle $.
\item[$*$] NNLO-QCD corrections to the dimension-six spectator contributions, $\tilde \Gamma_6^{(2)}$.
\item[$*$] Complete determination of LO-QCD dimension-seven contributions, 
$\Gamma_7^{(0)}$.
\end{itemize} 
As for the total decay rates, the HQE prediction is dominated by the free-quark decay. In this case, the theoretical uncertainties could be significantly reduced if the complete NNLO-QCD contributions, i.e.\ $\Gamma_3^{(2)}$, were available. Hence, the computation of the missing $\alpha_s^{(2)}$-corrections to non-leptonic $b$-quark decays is highly desirable.

In conclusion, combined with our recent studies on charmed hadrons \cite{Gratrex:2022xpm,King:2021xqp} and on $B$ mesons \cite{Lenz:2022rbq}, 
the results of this work confirm that the HQE provides a consistent framework to predict inclusive decay rates of heavy hadrons. 

\subsection*{Acknowledgements}
We would like to thank Olivier Schneider for providing us with old HFAG
averages for the $\Lambda_b$ lifetime. We would also like to thank Johannes Albrecht and Tim Gershon for useful correspondence on the prospects for measuring inclusive semileptonic decay rates of $b$-baryons at LHCb. We acknowledge support from the Alexander von Humboldt Foundation in the framework of the Research Group Linkage Programme, funded by the German Federal Ministry of Education. JG and BM have also been supported by the Croatian Science Foundation (HRZZ) project “Heavy hadron decays and lifetimes” IP-2019-04-7094. The work of MLP was funded by the Deutsche Forschungsgemeinschaft (DFG, German Research Foundation) - project number 500314741.
JG and IN wish to thank the theoretical particle physics group at the University of Siegen for the kind hospitality shown during their recent stay, where part of this work was undertaken.

\appendix

\section{Numerical inputs}
\label{app:1}
Here we collect the values of the parameters (table~\ref{tab:input}) and of the $\Delta B = 1$ Wilson coefficients (table~\ref{tab:WCs}) used in our analysis. 
\begin{table}[ht]
\small
\renewcommand{\arraystretch}{1.4}
    \centering
    \begin{tabular}{|c|c|c||c|c|c|}
    \hline
         Parameter & Value & Source & Parameter & Value & Source \\
    \hhline{|=|=|=||=|=|=|}
         $M_{B^+}$ & $5.27934 \GeV$ &
         \multirow{8}{*}{\cite{Workman:2022ynf}} &
         $|V_{us}|$  & $0.22500^{+0.00024}_{-0.00021}$ &
         \multirow{4}{*}{\cite{Charles:2004jd}}
         \\
         $M_{B_d}$ & $5.27965 \GeV$ &  & 
         $\displaystyle\frac{|V_{ub}|}{|V_{cb}|}$  & $0.08848^{+0.00224}_{-0.00219}$ & 
         \\
         $M_{B_s}$ & $5.36688 \GeV$ &  &
         $V_{cb}$  & $0.04145^{+0.00035}_{-0.00061}$ & 
         \\
         $M_{\Lambda_b}$ & $5.61960 \GeV$ & &
         $\delta $  & $\left(65.5^{+1.3}_{-1.2}\right)^\circ$ &
         \\
         \cline{4-6}
         $M_{\Xibm} $ & $5.7970 \GeV$ & &
         $m_b^{\rm kin}$ 
         & 
         $(4.573 \pm 0.012) \GeV$ & \cite{Bordone:2021oof} \\
         $M_{\Xibo} $ & $5.7919 \GeV$ & & 
         $\bar{m}_c (\bar{m}_c)$ & $ (1.27 \pm 0.02) \GeV$ 
         & \cite{Workman:2022ynf} 
         \\
          \cline{4-6}
         $M_{\Omega_b}$ & $6.9452 \GeV$ & &
         $f_B$ & $(0.1900 \pm 0.0013) \GeV$  &
         \multirow{2}{*}{\cite{Aoki:2019cca}} 
         \\
         $\alpha_s (M_Z)$  &  $0.1179 \pm 0.0010$ & &
         $f_{B_s}$ & $(0.2303 \pm 0.0013) \GeV$  & 
         \\
         \hline
    \end{tabular}
    \caption{Summary of inputs used in the numerical analysis.
    Values of the non-perturbative parameters for $b$-baryons are presented in tables~\ref{Tab:O1Numerics} and~\ref{Tab:parametrs2}. }
    \label{tab:input}
\end{table}
\begin{table}[ht]
\small
		\renewcommand{\arraystretch}{1.25}
		\centering
		\begin{tabular}
{|C{1.8cm}||C{1.8cm}|C{1.8cm}|C{1.8cm}|C{1.8cm}|C{1.8cm}|C{1.8cm}|}
			\hline 
			$\mu _1\text{[GeV]}$ & 2.5 & 4.2 & 4.5 & 4.8 & 9 \\
			\hline
			\multirow{2}{*}{$C_1 (\mu_1)$} 
			& 1.13 & 1.08 & 1.08 & 1.07 & 1.04 \\
			& (1.17) & (1.12) & (1.11) & (1.11) & (1.07) \\
			\hline
			\multirow{2}{*}{$C_2 (\mu_1)$} 
			& $-$0.27 & $-$0.19 & $-$0.18 & $-$0.17 & $-$0.11 \\
			& ($-$0.36) & ($-$0.27) & ($-$0.26) & ($-$0.25) & ($-$0.17) \\
			\hline
			\multirow{2}{*}{$C_3 (\mu_1)$} 
			& 0.02 & 0.01 & 0.01 & 0.01 & 0.01 \\
			& (0.02) & (0.01) & (0.01) & (0.01) & (0.01) \\
			\hline
			\multirow{2}{*}{$C_4 (\mu_1)$}
			& $-$0.05 & $-$0.04 & $-$0.03 & $-$0.03 & $-$0.02 \\
			& ($-$0.04) & ($-$0.03) & ($-$0.03) & ($-$0.03) & ($-$0.02) \\
			\hline
			\multirow{2}{*}{$C_5 (\mu_1)$}
			& 0.01 & 0.01 & 0.01 & 0.01 & 0.01 \\
			& (0.01) & (0.01) & (0.01) & (0.01) & (0.01) \\
			\hline
			\multirow{2}{*}{$C_6 (\mu_1)$}
			& $-$0.06 & $-$0.04 & $-$0.04 & $-$0.04 & $-$0.03 \\
			& ($-$0.05) & ($-$0.03) & ($-$0.03) & ($-$0.03) & ($-$0.02) \\
			\hline
			$C_8^{\rm eff} (\mu_1)$ 
			& ($-$0.17) & ($-$0.15) & ($-$0.15) & ($-$0.15) & ($-$0.14) \\
			\hline
		\end{tabular} 
		\caption{Values of the Wilson coefficients 
			at NLO(LO)-QCD for different choices of~$\mu_1$.}
		\label{tab:WCs}
	\end{table}
 
\section{Dimension-six four-quark operator contributions at LO-QCD}
\label{app:2}
The analytical expressions for the functions $\tilde \Gamma_{6,T}^q (x_{f_1},x_{f_2})$, introduced in eq.~\eqref{eq:dim-6-4q-NLO-scheme}, are provided explicitly below
at LO-QCD. For non-leptonic transitions $b \to q_1 \bar q_2 q_3$, with $q_{1,2} = u,c$ and $q_3 = d,s$, they read respectively 
\begin{align}
\tilde \Gamma_{6,{\rm int}^-}^{q_3} (x_{q_1}, x_{q_2}) &  = \frac{G_F^2}{12 \pi} |V_{q_1b}|^2 |V_{q_2 q_3}|^2 m_b^2 \sqrt{\lambda(1, x_{q_1}, x_{q_2})} \,
\Bigg\{ k_1 \Big[ \, \omega_1(x_{q_1},x_{q_2})\, \langle \Opsixprime{1}{q_3} \rangle
 - 2  \omega_2(x_{q_1}, x_{q_2}) \langle \Opsixprime{2}{q_3}\rangle  \Big]
 \nonumber \\[2mm]
& + k_2 \Big[
 \omega_1(x_{q_1}, x_{q_2}) \langle \Opsixtprime{1}{q_3} \rangle
- 2  \omega_2(x_{q_1}, x_{q_2}) \langle \Opsixtprime{2}{q_3}  
\rangle \Big] \Bigg\}\,,
\label{eq:Im-int-m}
\\[5mm]
\tilde \Gamma_{6,{\rm exc}}^{q_2} (x_{q_1}, x_{q_3}) & =  \frac{G_F^2}{2 \pi}  |V_{q_1b}|^2 |V_{q_2 q_3}|^2 m_b^2  \sqrt{\lambda(1, x_{q_1}, x_{q_3})}  \, (1 - x_{q_1} - x_{q_3}) \, \Big[ k_3 \langle \Opsixprime{1}{q_2} \rangle + k_4 \langle \Opsixtprime{1}{q_2}\rangle \Big]
\,,
\label{eq:ImT-exc}
\\[5mm]
\tilde \Gamma_{6,{\rm int}^+}^{q_1} (x_{q_2}, x_{q_3}) &  = \frac{G_F^2}{12 \pi} |V_{q_1b}|^2 |V_{q_2 q_3}|^2 m_b^2 \sqrt{\lambda(1, x_{q_3}, x_{q_2})} \,
\Bigg\{ k_5 \Big[ \, \omega_1(x_{q_3},x_{q_2})\, \langle \Opsixprime{1}{q_1} \rangle
 - 2  \omega_2(x_{q_3}, x_{q_2})\langle \Opsixprime{2}{q_1} \rangle  \Big]
 \nonumber \\[2mm]
& + k_6 \Big[
 \omega_1(x_{q_3}, x_{q_2}) \langle \Opsixtprime{1}{q_1}\rangle
- 2  \omega_2(x_{q_3}, x_{q_2}) \langle \Opsixtprime{2}{q_1} \rangle \Big] \Bigg\}\,,
\label{eq:Im-int-p}
\end{align}
while for semileptonic transitions $b \to q_1 \bar \nu_\ell \ell$, with $q_1 = u,c$ and $\ell = e, \mu, \tau$, the explicit expression is
\begin{align}
\tilde \Gamma_{6,{\rm int}^+}^{q_1} (x_{\ell}, x_{\nu_\ell}) &  = \frac{G_F^2}{12 \pi} |V_{q_1b}|^2 m_b^2 \sqrt{\lambda(1, x_\ell, x_{\nu_\ell})} \,
\Big[ \, \omega_1(x_\ell, x_{\nu_\ell})\, \langle  \Opsixprime{1}{q_1} \rangle
 - 2  \omega_2(x_\ell, x_{\nu_\ell})\langle \Opsixprime{2}{q_1} \rangle  \Big] \,,
\label{eq:Im-int-p-sl}
\end{align}
where $x_{f} = m_{f}^2/m_b^2$ and $\lambda(a,b,c) =(a - b - c)^2 - 4 b c $ is the K\"allen function. Moreover, in eqs.~\eqref{eq:Im-int-m}-\eqref{eq:Im-int-p-sl} we have introduced the functions $\omega_{1,2}(a,b)$, symmetric in their arguments, with
\begin{align}
\omega_1(a,b)  = (a - b)^2 + a + b - 2\,,
\qquad
\omega_2(a,b) = 2 \, (a -b)^2 - (1 + a + b)\,,
\end{align}
while $k_{1}, \ldots, k_6,$ denote the following combinations of Wilson coefficients:
\begin{alignat}{4}
& k_1 =  2 \, C_1 C_2 + N_c \, C_2^2 \,, \qquad 
&& k_2 = \, C_1^2\,,
\\
& k_3 = 2 \, C_1 C_2 \,, \qquad
&& k_4 = \, \Big( C_1^2 + C_2^2 \Big) \,,
\\
& k_5 =  N_c  \, C_1^2 + 2 \, C_1 C_2 \,, \qquad 
&& k_6 = \, C_2^2\,.
\label{eq:WC-WE}
\end{alignat}

\bibliographystyle{JHEP}
\bibliography{References}

\end{document}